\documentclass[aps,prl,twocolumn,superscriptaddress,floatfix]{revtex4-2}

\usepackage{tabularx}
\usepackage{graphicx}
\usepackage{dcolumn}
\usepackage{bm, color}
\usepackage{amsmath}
\usepackage{amsfonts}
\usepackage{amssymb}
\usepackage{amsmath}
\usepackage{here}
\usepackage{multirow}
\usepackage{ulem}
\usepackage{bbm}
\usepackage{cancel}
\providecommand{\U}[1]{\protect\rule{.1in}{.1in}}

\usepackage{hyperref}
\hypersetup{
	colorlinks = true,
}
\usepackage{physics}

\usepackage[whole]{bxcjkjatype}

\begin{document}

\title{Probing the Dirac Quantum Geometry in Quantum Magnets from Polarized Raman Spectroscopy}

\author{Yuki Yamazaki}
\affiliation{Department of Physics, Graduate School of Science, Kyoto University, Kyoto 606-8502, Japan}
\author{Doh Young Kim}
\affiliation{Department of Physics and Astronomy, Seoul National University, Seoul 08826, Korea}
\affiliation{Center for Theoretical Physics (CTP), Seoul National University, Seoul 08826, Korea}
\affiliation{Institute of Applied Physics, Seoul National University, Seoul 08826, Korea}
\author{Takahiro Morimoto}
\email{morimoto@scphys.kyoto-u.ac.jp}
\affiliation{Department of Physics, Graduate School of Science, Kyoto University, Kyoto 606-8502, Japan}
\author{Bohm-Jung Yang}
\email{bjyang@snu.ac.kr}
\affiliation{Department of Physics and Astronomy, Seoul National University, Seoul 08826, Korea}
\affiliation{Center for Theoretical Physics (CTP), Seoul National University, Seoul 08826, Korea}
\affiliation{Institute of Applied Physics, Seoul National University, Seoul 08826, Korea}
\date{\today}

\begin{abstract}
Magnetic Raman spectroscopy is sensitive to quasiparticle quantum geometry, yet microscopic Raman vertices generally prevent a direct geometric interpretation. Here we establish a symmetry-based extraction principle for two-dimensional quantum magnets with two-band quasiparticle excitations. Opposite circular-polarization spectra separate vertex-weighted quantum-metric and Berry-curvature contractions. Near a $C_3$-symmetric massive Dirac valley, symmetry fixes the leading Raman and velocity doublets up to a common scale, allowing the unknown Raman strength and intensity normalization to cancel from their spectral ratio. For an inversion-related pair of massive Dirac cones, the normalized counter-circular Raman asymmetry at frequency threshold $\Omega_{\rm th}$ is fixed by the Chern number and follows a universal near-threshold frequency scaling. This yields a local spectroscopic readout without absolute-intensity calibration or frequency integration. We demonstrate the principle for Majorana fermions in the field-induced honeycomb Kitaev spin liquid and for topological magnons in a honeycomb ferromagnet, and identify the conditions for resolving it experimentally.
\end{abstract}

\maketitle

\begin{figure}[t]
\centering
\includegraphics[width=\linewidth]{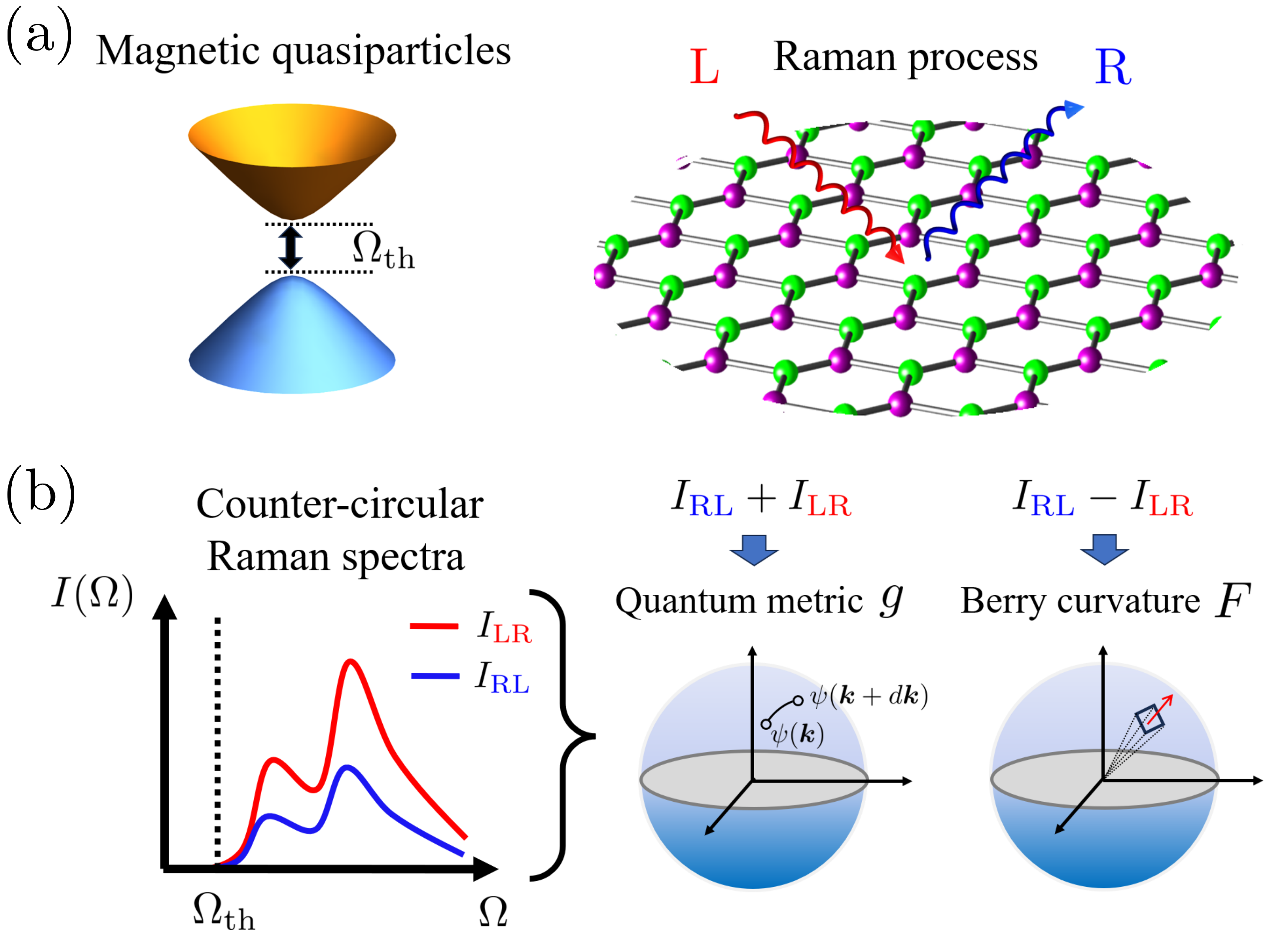}
\caption{Quantum-geometric Raman spectroscopy of magnetic quasiparticles. (a) Schematic massive-Dirac quasiparticle spectrum with Raman
threshold $\Omega_{\rm th}$ and polarization-resolved Raman scattering from a magnetic system. Here $L/R$ denote left-/right-circular polarizations defined with respect to the same oriented sample frame $(X,Y)$ for both incoming and outgoing photons. (b) Counter-circular Raman spectra $I_{RL}(\Omega)$ and $I_{LR}(\Omega)$, where the first (second) label denotes the incoming (outgoing) photon polarization. Their spectral sum $I_{RL}+I_{LR}$ probes an energy-resolved quantum-metric contraction, whereas their difference $I_{RL}-I_{LR}$ probes a Berry-curvature contraction. In the $C_3$-symmetric massive-Dirac regime, the normalized counter-circular asymmetry directly determines the Chern number at threshold, with a universal leading-Dirac scaling form near threshold.}
\label{fig}
\end{figure}

\textit{Introduction.—}
The quantum geometric tensor (QGT) describes the wave-function structure of quasiparticle bands beyond their energy dispersion. Its real and imaginary parts encode the quantum metric and Berry curvature, respectively, underlying a broad range of geometric and topological responses~\cite{Rhim59, Hwang6433, Jung035134, Tianyu334,Yu101,Vermaa226,Anyuan2026,Yang_2026}. Recent photoemission experiments have made electronic wave-function geometry accessible in solids~\cite{Kang110,Kim1050}, opening the prospect of characterizing quantum materials through both their dispersions and their eigenstates~\cite{Yang_2026}.

Extending this program to magnetic quasiparticles presents a distinct challenge. Magnons and fractionalized Majorana excitations are electrically neutral and require probes different from those used for electronic Bloch bands. Nevertheless, experiments have revealed magnetic excitations whose wave-function geometry is of considerable interest. In the Kitaev candidate $\alpha$-RuCl$_3$, Raman and neutron-scattering continua provide evidence consistent with fractionalization~\cite{Sandilands147201,Do2017NatPhys,Banerjee4604}, while reports of half-quantized thermal Hall conductance have motivated a field-induced topological Majorana interpretation~\cite{Kasahara559,Yokoi568}. The interpretation remains debated~\cite{Matsuda045003}, making complementary spectroscopic tests valuable. Magnon Hall transport and neutron-scattering observations of topological magnon bands likewise motivate spectroscopic probes of magnetic band geometry~\cite{Onose2010Science,Chisnell147201,Chen041028}. Transport and dispersion measurements provide important evidence, but do not by themselves determine the quasiparticle QGT.

Several routes to magnetic quantum geometry have recently emerged. Momentum-resolved mode tomography from dynamical structure factors has recently been proposed to reconstruct the QGT of bosonic one-particle excitations~\cite{Wu2026}. 
This approach, however, relies on resolving the momentum-space eigenmode structure and is not directly applicable to continua generated by fractionalized or multiparticle excitations.

Recent work has established the quantum geometric content of magnetic Raman circular dichroism (RCD). Bostr\"om et al. related honeycomb-magnet RCD to magnon Berry curvature and showed that its frequency-integrated response can track a Chern-phase diagram~\cite{Bostrom026701}. Koller et al.~\cite{Koller194425} showed that spin-liquid dichroism contains broader wave-function-geometric contributions. Liu et al.~\cite{Liu2026} developed a microscopic light--matter-coupling (LMC) formulation in which the two-band magnon response is Berry curvature multiplied by a momentum-dependent geometric weight. Quantum-metric signatures in Raman spectra have also been discussed for altermagnetic magnons~\cite{Yuan2026} and chiral phonons in a quantum magnet~\cite{Wang2026RamanMetric}.

These developments establish Raman spectroscopy as a geometry-sensitive probe, but leave an extraction problem: the geometric response generally remains weighted by microscopic Raman vertices. A dichroic spectrum alone consequently does not identify a Chern number, and its absolute magnitude depends on the LMC. The questions we should address are whether polarization channels can separate metric and curvature contributions, whether symmetry can eliminate the unknown vertex weights, and whether the resulting observable can identify topology directly at a resolved Raman shift, without integrating the spectrum.

In this Letter, we establish such a symmetry-based extraction principle for two-band Dirac quasiparticles in two-dimensional quantum magnets. First, for a symmetric Hermitian Raman tensor, the counter-circular spectral sum and difference separate energy-resolved metric and curvature contractions. Their generic weights retain microscopic information. The central advance is that $C_3$ symmetry constrains the leading symmetric-traceless Raman doublet relative to the Dirac velocity doublet up to a common scale and an orthogonal transformation. 
The scale then cancels from the normalized counter-circular response. 
For an inversion-related valley pair adiabatically connected to a two-cone parent without additional topological contributions, the normalized counter-circular Raman asymmetry at frequency threshold $\Omega_{\rm th}$ is fixed by the Chern number and follows a universal near-threshold frequency scaling. This yields a local spectroscopic readout without absolute-intensity calibration or frequency integration.
We demonstrate this mechanism for Majorana fermions in a field-induced Kitaev spin liquid on the honeycomb lattice and for topological magnons in a number-conserving honeycomb ferromagnet.
The result provides a common geometric readout for two physically different Raman processes, with a clear distinction between local threshold selection and its global topological interpretation.


\textit{Raman--QGT correspondence.—}
We consider Raman scattering in a translationally invariant quasiparticle description and work in the zero-momentum-transfer limit. An incident photon of frequency $\omega_{\rm in}$ is scattered into an outgoing photon of frequency $\omega_{\rm out}$, transferring the Raman shift $\Omega\equiv\omega_{\rm in}-\omega_{\rm out}>0$ to the system. For a polarization geometry $P=(\bm e_{\rm in},\bm e_{\rm out})$, we represent the corresponding low-energy Raman process by an effective quasiparticle vertex $R_P(\bm k)$. For definiteness, we first write the zero-temperature Raman response as
\begin{align}
\hat R_P &= \sum_{\bm k}\Psi_{\bm k}^{\dagger}R_P(\bm k)\Psi_{\bm k},
\nonumber\\
I_P(\Omega) &= {\cal N}_R \sum_f\left|\langle f|\hat R_P|0\rangle\right|^2\delta\!\left[\Omega-(E_f-E_0)\right].
\label{eq:PRL_Raman_start}
\end{align}
Here $\Psi_{\bm k}$ is the effective quasiparticle spinor, $|0\rangle$ is the ground state with energy $E_0$, $|f\rangle$ is an excited many-body state with energy $E_f$, and ${\cal N}_R$ collects common polarization-independent normalization factors. The finite-temperature generalization amounts to a thermal average over initial many-body states and will be used explicitly for the magnon response below. For a quasiparticle transition $|u_n(\bm k)\rangle\to|u_m(\bm k)\rangle$, the corresponding band-basis Raman amplitude is $M_P^{mn}(\bm k) =\langle u_m(\bm k)|R_P(\bm k)|u_n(\bm k)\rangle$.

The effective nonresonant Loudon--Fleury (LF)~\cite{Fleury514, Shastry1068} and LMC~\cite{Bostrom026701, Koller194425, Liu2026} Raman vertices considered below have Hermitian Cartesian components and a symmetric Raman tensor. Our Raman–QGT correspondence is formulated for this class of vertices. More general magnetic Raman processes, including resonant processes with complex vertices, as well as anomalous bosonic Bogoliubov Raman processes, require a separate treatment.
We specialize to a two-dimensional two-band Hamiltonian $h(\bm k)=d_0(\bm k)\sigma_0+\bm d(\bm k)\cdot\bm\sigma$, with nondegenerate eigenstates $|u_\pm(\bm k)\rangle$, eigenenergies $E_\pm(\bm k)$, and band separation $\Delta(\bm k)=E_+(\bm k)-E_-(\bm k)>0$. We use $i,j=x,y$ for an oriented Cartesian momentum frame aligned with the physical polarization axes $X,Y$. Defining $V_i(\bm k) =\langle u_+(\bm k)|\partial_{k_i}h(\bm k)|u_-(\bm k)\rangle$, the lower-band QGT is $Q_{ij} =\langle\partial_{k_i}u_-|(1-|u_-\rangle\langle u_-|) |\partial_{k_j}u_-\rangle =V_i^*V_j/\Delta^2 =g_{ij}-iF_{ij}/2$, where $g_{ij}$ and $F_{ij}$ indicate the quantum metric and Berry curvature, respectively.

At fixed $\bm k$, the two quantities $V_x$ and $V_y$ are complex numbers. If ${\rm Im}(V_x^*V_y)\neq0$, they are linearly independent over $\mathbb R$ and therefore form a basis of $\mathbb C$ regarded as a two-dimensional real vector space. Hence, for any Hermitian Raman component $R_\mu(\bm k)$ with $M_\mu=\langle u_+|R_\mu|u_-\rangle$, there exist unique real coefficients $\lambda_\mu^i(\bm k)$ such that $M_\mu=\lambda_\mu^iV_i$, explicitly
$\lambda_\mu^x={\rm Im}(M_\mu^*V_y)/{\rm Im}(V_x^*V_y)$ and $\lambda_\mu^y={\rm Im}(V_x^*M_\mu)/{\rm Im}(V_x^*V_y)$. Using the QGT relation above,
${\rm Im}(V_x^*V_y)=-\Delta^2F_{xy}/2$. Thus, the real velocity basis is nonsingular wherever $F_{xy}\neq0$. For an energy-resolved response, however, it is sufficient that the set on which this decomposition fails have zero measure on each regular constant-$\Delta$ Raman shell.

For the symmetric Raman tensors considered below, $R_{XY}=R_{YX}$. We define $R_B=R_{XX}-R_{YY}$ and $R_C=R_{XY}+R_{YX}$. With $\bm e_{R/L}=(\hat{\bm X}\pm i\hat{\bm Y})/\sqrt2$, we label the first (second) polarization index by the incoming (outgoing) photon and use $R_{s_{\rm in}s_{\rm out}} =\bm e_{s_{\rm out}}^\dagger R\bm e_{s_{\rm in}}$. Then $2R_{RL}=R_B+iR_C$ and $2R_{LR}=R_B-iR_C$. Since the common factor of $1/2$ produces only a common polarization-independent intensity factor, we absorb it into ${\cal N}_R$ and write the corresponding interband amplitudes as $M_{RL}=M_B+iM_C$ and $M_{LR}=M_B-iM_C$.
Their spectra are $I_{s}(\Omega) ={\cal N}_R\int_{\bm k}|M_s(\bm k)|^2 \delta[\Omega-\Delta(\bm k)]$ for $s=RL,LR$, where $\int_{\bm k}\equiv\int d^2k/(2\pi)^2$.
We define $I_\Sigma=I_{RL}+I_{LR}$ and $I_\Delta=I_{RL}-I_{LR}$. Using $M_{B,C}=\lambda_{B,C}^iV_i$ and the two-band QGT identity, we obtain
\begin{align}
\frac{I_\Sigma(\Omega)}{\Omega^2} &= 2{\cal N}_R \int_{\bm k} {\cal G}^{ij}_\Sigma(\bm k)g_{ij}(\bm k)\delta[\Omega-\Delta(\bm k)],
\nonumber\\
\frac{I_\Delta(\Omega)}{\Omega^2} &= 2{\cal N}_R \int_{\bm k} {\cal F}_\Delta(\bm k)F_{xy}(\bm k)\delta[\Omega-\Delta(\bm k)],
\label{eq:PRL_QGT_spectra}
\end{align}
where ${\cal G}^{ij}_\Sigma \equiv \lambda_B^i\lambda_B^j+\lambda_C^i\lambda_C^j$ and ${\cal F}_\Delta \equiv \lambda_B^x\lambda_C^y-\lambda_B^y\lambda_C^x$. Summation over repeated indices $i,j=x,y$ is understood in the first line.
Thus, the counter-circular sum and difference isolate energy-resolved quantum-metric and Berry-curvature contractions, respectively. Equation~\eqref{eq:PRL_QGT_spectra} is an exact representation of the two-band interband Raman response. It does not, by itself, constitute a reconstruction of the momentum-resolved QGT, since the coefficients $\lambda_\mu^i(\bm k)$ generally retain microscopic information about the quasiparticle Hamiltonian and Raman vertex. For a symmetric Raman tensor, the co-circular channels satisfy $R_{RR}=R_{LL}$, so their dichroic difference vanishes identically, whereas linear polarizations involve only real combinations of Raman components and therefore probe only quantum-metric contractions. The counter-circular pair is thus distinguished in separating the quantum-metric and Berry-curvature sectors through its spectral sum and difference.

\textit{Symmetry-constrained massive-Dirac response.—}
While the coefficients $\lambda_\mu^i(\bm k)$ remain model-dependent, this microscopic freedom is strongly reduced for $C_3$-symmetric massive Dirac systems. As a representative example, we consider a graphene-type two-sublattice  honeycomb model, where $C_3$ and sublattice (chiral) symmetries produce a gapless Dirac cone~\cite{RevModPhys.81.109, PhysRevB.90.115207}. Introducing a perturbation that breaks the chiral symmetry while preserving $C_3$ can lead to a $ C_3$-symmetric massive Dirac model. Such graphene-type low-energy structure is realized in the Kitaev Majorana and honeycomb-magnon models as considered below.

A $C_3$-symmetric massive Dirac structure implies the following relation between the Raman and velocity matrix elements. Once a $C_3$-symmetric massive Dirac doublet is formed, the in-plane velocity doublet $(v_x,v_y)$ and the symmetric-traceless Raman doublet $(R_B,R_C)$ transform equivalently under $C_3$. Consequently, symmetry fixes their leading relation up to a single overall scale and an orthogonal transformation. Choosing a positive-chirality local basis, we obtain
\begin{align}
h_K(\bm q) &= \epsilon_K\sigma_0 +v_K(q_x\sigma_x+q_y\sigma_y)+m_K\sigma_z+O(q^2),
\nonumber\\
\begin{pmatrix}
M_B^{(0)}\\
M_C^{(0)}
\end{pmatrix}
&= \lambda_K O_K
\begin{pmatrix}
V_x^{(0)}\\
V_y^{(0)}
\end{pmatrix},
\label{eq:PRL_C3_theorem}
\end{align}
where $h_K(\bm q)$ is the effective Hamiltonian near the $K$-valley with $\bm q=\bm k-\bm K$. $\lambda_K>0$, and $O_K$ is an orthogonal matrix satisfying $O_K^{\mathsf T}O_K=\mathbbm{1}_2$ and $\det O_K=-1$. Here $V_i^{(0)}$ and $M_{B,C}^{(0)}$ denote the leading ($O(q^0)$) interband velocity and Raman matrix elements, respectively. 
The Hamiltonian, Raman operators, and physical velocity vertices must be expressed in a compatible local representation, as specified in the Supplemental Materials (SM). The symmetry derivation is given in the End Matter.
Equation~\eqref{eq:PRL_C3_theorem} then gives $\mathcal G_\Sigma^{ij,K}=\lambda_K^2\delta_{ij}$ and $\mathcal F_\Delta^K=-\lambda_K^2$.
Thus, to leading order near the Dirac point, the otherwise model-dependent Raman coefficients reduce to a single overall scale.

We now consider two valleys $\bm K$ and $\bm K'$ exchanged by inversion. Inversion maps $\bm K+\bm q$ to $\bm K'-\bm q$ and enforces identical Dirac dispersions at the two valleys. Consequently, $v_K=v_{K'}\equiv v$ and $|m_K|=|m_{K'}|\equiv m$. Inversion covariance of the Raman tensor likewise gives equal positive Raman scales, $\lambda_K=\lambda_{K'}\equiv\lambda$. The two valleys therefore have the common threshold $\Omega_{\rm th}=2m$. Moreover, the quantum metric and Berry curvature at the two valleys are related by inversion, so their contributions to both $I_\Sigma$ and $I_\Delta$ add.

To distinguish the local Dirac information from the global band topology, we define the Dirac index $\nu_D \equiv \operatorname{sgn}F_{xy}^{K}(\bm q=\bm 0) = \operatorname{sgn}F_{xy}^{K'}(\bm q=\bm 0)=\pm1$, where the equality follows from inversion symmetry. Within the leading massive-Dirac theory, the Raman shell at frequency $\Omega$ has characteristic momentum $q_\Omega=\sqrt{\Omega^2-\Omega_{\rm th}^2}/(2v)$. Keeping the linear Dirac Hamiltonian and the momentum-independent Raman doublet fixed by $C_3$, the valley-summed counter-circular response takes the universal scaling form $\frac{I_\Delta(\Omega)}{I_\Sigma(\Omega)} =
-\nu_D \frac{2\Omega_{\rm th}\Omega}{\Omega^2+\Omega_{\rm th}^2}$. This finite-frequency scaling form is exact within the leading massive-Dirac theory and applies while $q_\Omega$ remains within the low-energy Dirac regime, where higher-order momentum corrections to the Dirac Hamiltonian and momentum-dependent Raman-vertex corrections are negligible.

For the lattice systems considered here, the Dirac mass term can be continuously reduced to the weak-mass regime without closing the bulk gap elsewhere. In this limit, the Berry curvature is localized near the two inversion-related Dirac valleys, each of which contributes $\nu_D/2$ to the lower-band Chern number. Hence $C=\nu_D=\pm1$, and this value is unchanged upon restoring the finite mass. Consequently, within the leading Dirac regime,
\begin{align}
\frac{I_\Delta(\Omega)}{I_\Sigma(\Omega)}
=
-C\,\frac{2\Omega_{\rm th}\Omega}
{\Omega^2+\Omega_{\rm th}^2}, \quad \lim_{\Omega\to\Omega_{\rm th}^{+}}
\frac{I_\Delta(\Omega)}{I_\Sigma(\Omega)}
=
-C.
\label{eq:PRL_Chern_ratio}
\end{align}
The threshold limit is more robust: provided that the symmetry-allowed zeroth-order Raman doublet is nonzero and that the Dirac valley pair gives the isolated lowest interband threshold, as $\Omega\to\Omega_{\rm th}^{+}$ the Raman shell collapses onto the Dirac valleys, so that smooth momentum-dependent corrections vanish from the normalized response.
The common Raman normalization ${\cal N}_R$, the microscopic Raman scale $\lambda$, and the Dirac velocity $v$ have all canceled. Equation~\eqref{eq:PRL_Chern_ratio} therefore determines the Chern number directly from the threshold counter-circular asymmetry, while its finite-frequency form gives the leading-Dirac scaling behavior near threshold. Neither result requires absolute-intensity calibration or frequency integration. We note that this cancellation concerns a common multiplicative intensity normalization; relative calibration between the $RL$ and $LR$ channels, polarization leakage, and additive backgrounds must still be controlled experimentally.

\textit{Majorana excitations in the honeycomb Kitaev spin liquid.—}
We now apply Eq.~\eqref{eq:PRL_Chern_ratio} to the field-induced Kitaev spin liquid. The Kitaev honeycomb model is
\begin{align}
H_K = -\sum_{\mu=x,y,z}J_\mu\sum_{\langle ij\rangle_\mu}S_i^\mu S_j^\mu ,
\label{eq:Kitaev_H}
\end{align}
where $\langle ij\rangle_\mu$ denotes a nearest-neighbor bond of type $\mu=x,y,z$. We focus on the isotropic point $J_x=J_y=J_z=J$. A weak magnetic field generates the leading flux-preserving three-spin interaction
\begin{align}
H_\eta = 2\eta \sum_{\langle ijk\rangle_{\alpha\beta}}S_i^\alpha S_j^\gamma S_k^\beta, \qquad \eta\propto h_xh_yh_z ,
\label{eq:PRL_Kitaev_three_spin}
\end{align}
where $\langle ijk\rangle_{\alpha\beta}$ is an oriented two-bond path and $(\alpha,\beta,\gamma)$ is a permutation of $(x,y,z)$. This term gaps the Majorana Dirac cones~\cite{Kitaev2,Motome012002}.

In the flux-free sector, the model reduces to a quadratic two-sublattice Majorana Hamiltonian $h_{\rm Maj}(\bm k)=d_x(\bm k)\sigma_x+d_y(\bm k)\sigma_y+d_z(\bm k)\sigma_z$. The nearest-neighbor terms generate the two gapless Dirac cones, whereas the field-induced three-spin term provides the mass component
$d_z(\bm k)$. The magnetic Raman vertex contains the nearest-neighbor LF contribution and, in a magnetic field, a contribution associated with the three-spin term. For the symmetric-traceless channels $P=B,C$, we write $R_P^{\rm Maj} =R_{{\rm LF},P}^{\rm Maj}+\zeta_R R_{\eta,P}^{\rm Maj}$. Following the field-induced LMC construction of Ref.~\cite{Koller194425}, $R_{\eta,\alpha\beta}^{\rm Maj}(\bm k) =-\partial_{k_\alpha}\partial_{k_\beta} d_z(\bm k)\sigma_z$.
The parameter $\zeta_R$ parametrizes the microscopic coupling of this field-induced term to light and is not fixed by $\eta$ alone. The explicit functions $d_a(\bm k)$ and $R_P^{\rm Maj}$ are given in the SM~\cite{SM}. After accounting for the Majorana particle-hole redundancy, the positive-frequency pair-creation amplitude is represented by the same negative-to-positive-energy interband matrix element $\langle u_+(\bm k)|R_P^{\rm Maj}(\bm k)|u_-(\bm k)\rangle$ entering Eq.~\eqref{eq:PRL_QGT_spectra}.

At the isotropic point, the flux-free Majorana Hamiltonian has two inversion-related massive Dirac valleys at $\bm K$ and $\bm K'=-\bm K$. 
In the local isotropized momentum frame, the Hamiltonian takes the form $h_\tau(\bm\kappa)
=(J/2)\kappa_x\sigma_x +(J/2)\chi_\tau\kappa_y\sigma_y-\chi_\tau(3\sqrt3/2)\eta\,\sigma_z$, where $\tau=K,K'$ and $\chi_K=+1$, $\chi_{K'}=-1$. After transforming the velocity vertices back to the common oriented physical Cartesian frame, the corresponding leading symmetric-traceless Raman amplitudes satisfy $(M_{B,\tau}^{(0)},M_{C,\tau}^{(0)})^{\mathsf T} =(1/\sqrt3)\operatorname{diag}(\chi_\tau,-\chi_\tau)(V_{X,\tau}^{(0)},V_{Y,\tau}^{(0)})^{\mathsf T}$. Here $V_{X,\tau}^{(0)}$ and $V_{Y,\tau}^{(0)}$ are the leading interband velocity matrix elements defined with respect to the same oriented physical Cartesian frame used above. Thus, both valleys satisfy Eq.~\eqref{eq:PRL_C3_theorem} with the common scale $\lambda_K=\lambda_{K'}=1/\sqrt3$ and an orthogonal transformation of determinant $-1$. Moreover, the field-induced Raman components $R_{\eta,B}^{\rm Maj}$ and $R_{\eta,C}^{\rm Maj}$ vanish at the valley centers. Hence, Eq.~\eqref{eq:PRL_C3_theorem} and the corresponding normalized threshold response are independent of $\zeta_R$.
With the orientation convention specified in the SM~\cite{SM}, the two Dirac masses are $m_\tau=-\chi_\tau 3\sqrt3\,\eta/2$. The resulting Raman threshold and lower-band Majorana Chern number are $\Omega_{\rm th}=3\sqrt3|\eta|$ and $C=\operatorname{sgn}(\eta)$.

\begin{figure}[t]
\centering
\includegraphics[width=\linewidth]{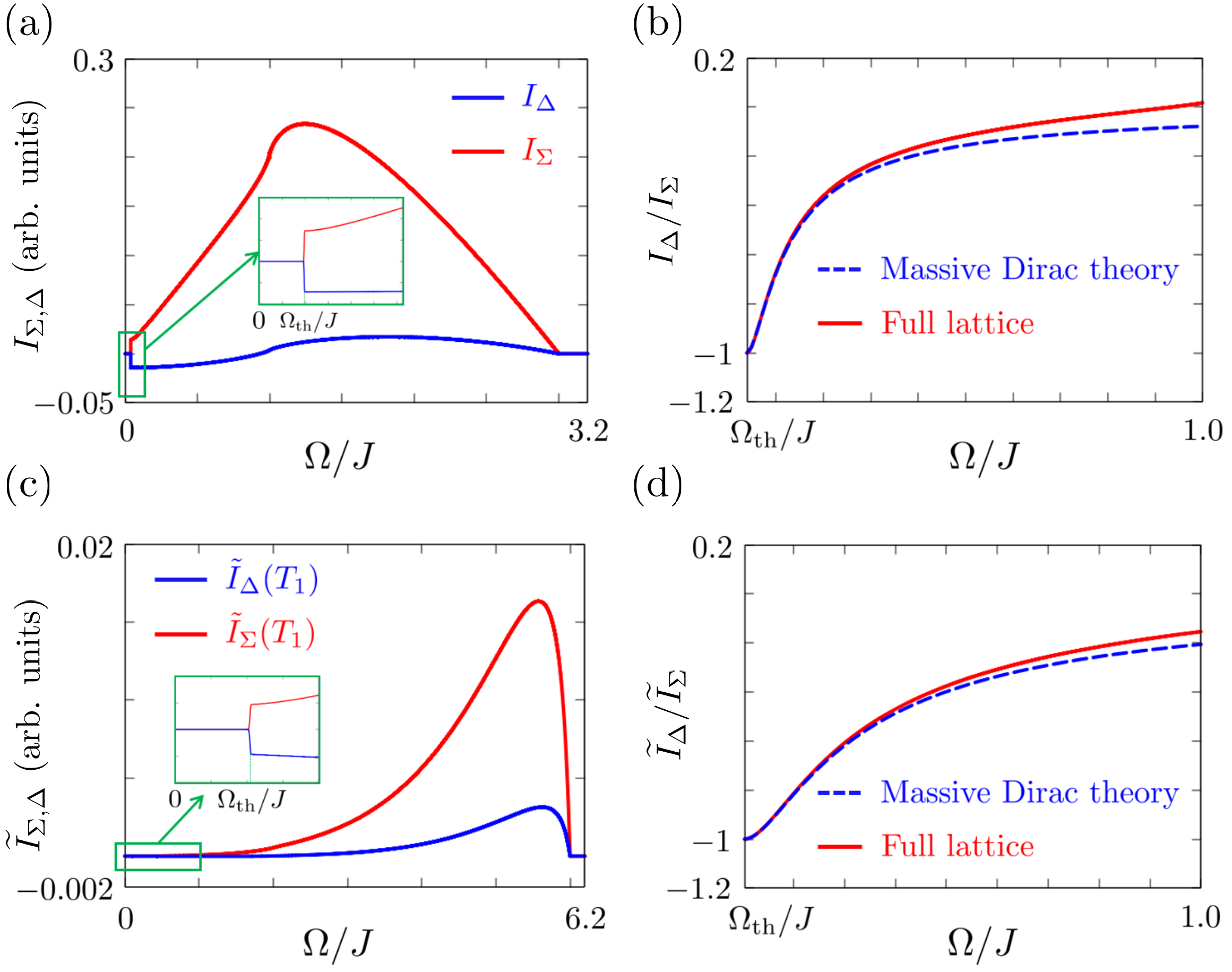}
\caption{Full-lattice calculation of Raman signatures of Majorana and magnon quantum geometry on the honeycomb lattice. (a) Majorana counter-circular sum $I_\Sigma$ and difference $I_\Delta$ for $\eta/J=0.0075$ and $\zeta_R=1$ with $\Omega_{\rm th}/J\sim 0.039$. (b) Normalized counter-circular asymmetry $I_\Delta/I_\Sigma$, compared with the leading massive-Dirac prediction in Eq.~\eqref{eq:PRL_Chern_ratio} plotted for $\Omega/J\in[\Omega_{\rm th}/J,1]$. (c) Topological-magnon sum $\widetilde I_\Sigma$ and difference $\widetilde I_\Delta$ at $T_1/J=0.4$ for $D/J=0.01$, $B_z/J=0.1$, and $S=1$ with $\Omega_{\rm th}/J\sim 0.1$. (d) Normalized magnon asymmetry $\widetilde I_\Delta/\widetilde I_\Sigma$. The magnon intensities depend on temperature, whereas the Bose occupation factor cancels from their normalized ratio. In both systems, the asymmetry approaches minus the lower-band Chern number near the respective Raman thresholds $\Omega_{\rm th}/J$, with higher-frequency deviations arising from lattice corrections.}
\label{fig:lattice_PRL}
\end{figure}

We test the leading-Dirac relation and its universal threshold limit using the full-lattice Majorana Hamiltonian and the full momentum-dependent Raman vertices of this effective flux-free model. Figures~\ref{fig:lattice_PRL}(a,b) show the full-lattice result for $\eta/J=0.0075$ and $\zeta_R=1$. Figure~\ref{fig:lattice_PRL}(b) compares the normalized counter-circular asymmetry $I_\Delta/I_\Sigma$ with the leading massive-Dirac prediction in Eq.~\eqref{eq:PRL_Chern_ratio}. The full-lattice result approaches $-C=-1$ at threshold and follows the Dirac scaling near threshold, with deviations appearing as the Raman shell leaves the leading Dirac regime.

\textit{Topological magnon system.—}
Another quantum magnetic system hosting Dirac-type quasiparticle excitations is a number-conserving honeycomb ferromagnet with nearest-neighbor exchange $J$, next-nearest-neighbor Dzyaloshinskii--Moriya interaction $D$, and an out-of-plane magnetic field $B_z$~\cite{Owerre2016,Kim227201,Chen041028,Chen031047}:
\begin{align}
\nonumber
\hat H_{\rm FM} = -J\sum_{\langle ij\rangle}\bm S_i\cdot\bm S_j &-D\sum_{\langle\langle ij\rangle\rangle}\nu_{ij}\, \hat{\bm z}\cdot(\bm S_i\times\bm S_j)
\\&-B_z\sum_i S_i^z .
\label{eq:magnon_spin_H}
\end{align}
Here $\nu_{ij}=-\nu_{ji}=\pm1$ specifies the orientation of the next-nearest-neighbor bonds; our convention is chosen such that $D>0$ gives Chern number $C_{\rm M}=+1$ for the lower magnon band. For $B_z>0$, we expand about the ferromagnetic state polarized along $+\hat{\bm z}$. Linear spin-wave theory then yields a number-conserving Hermitian two-band Bloch Hamiltonian $h_{\rm mag}(\bm k)$, so the conventional band QGT applies directly.

For this minimal model, the nontrivial Pauli-vector part of $h_{\rm mag}$ is identical to that of the Majorana Hamiltonian under the replacements $J\to 2JS$ and $\eta\to2DS$, while the magnon Hamiltonian additionally contains the scalar shift $\varepsilon_D=3JS+B_z$. Accordingly, the two magnon bands are $\epsilon_\pm(\bm k)=\varepsilon_D\pm E(\bm k)$, with threshold $\Omega_{\rm th}=6\sqrt3|D|S$ and lower-band Chern number $C_{\rm M}=\operatorname{sgn}(D)$.

For the LMC Raman vertex adopted here, the same mapping holds for the symmetric-traceless Raman components: $R_{{\rm LF},P}^{\rm mag}=R_{{\rm LF},P}^{\rm Maj}|_{J\to 2JS}$ and $R_{{\rm DMI},P}^{\rm mag} =R_{\eta,P}^{\rm Maj}|_{\eta\to2DS}$, with $P=B,C$. Thus, within this minimal honeycomb model, the full magnon Raman vertex maps onto the Majorana vertex with $\zeta_R=1$. The microscopic derivation and sign conventions are given in the SM~\cite{SM}.

The physical scattering process, however, is different. Whereas the Majorana response corresponds to pair creation, the positive-frequency number-conserving magnon Stokes process promotes a thermally occupied magnon from the lower to the upper band at the same momentum. Writing $\alpha_{\pm,\bm k}$ for the corresponding band operators, the relevant term is $\hat R^{\rm inter}_{P,\mathrm{Stokes}} =\sum_{\bm k}M_P(\bm k)\alpha^\dagger_{+,\bm k}\alpha_{-,\bm k}$.
The associated finite-temperature scattering intensity is weighted by $B_T(\bm k) =n_B[\epsilon_-(\bm k)]\{1+n_B[\epsilon_+(\bm k)]\}$, where $n_B(\epsilon)=(e^{\epsilon/T}-1)^{-1}$ and $k_B=1$.

On the Raman shell, $\Omega=\epsilon_+(\bm k)-\epsilon_-(\bm k)=2E(\bm k)$. Because the scalar part $\varepsilon_D$ is momentum independent in the present model, this condition fixes the two band energies separately as $\epsilon_\pm=\varepsilon_D\pm\Omega/2$. Hence, the occupation factor is constant over each Raman shell,
$B_T(\bm k)=B_T(\Omega)$, with $B_T(\Omega)=n_B(\varepsilon_D-\Omega/2)[1+n_B(\varepsilon_D+\Omega/2)]$. The finite-temperature spectra therefore satisfy $\widetilde I_{\Sigma,\Delta}(\Omega,T) =B_T(\Omega)I_{\Sigma,\Delta}(\Omega)$, and the thermal factor cancels exactly from the normalized response: $\widetilde I_\Delta/\widetilde I_\Sigma =I_\Delta/I_\Sigma$. The leading massive-Dirac relation in Eq.~\eqref{eq:PRL_Chern_ratio} consequently applies with $C\to C_{\rm M}$.

Figures~\ref{fig:lattice_PRL}(c,d) show the full-lattice analysis result. Although the Raman intensity depends on temperature, the normalized counter-circular asymmetry $\widetilde I_\Delta/\widetilde I_\Sigma$ is independent of the common Bose occupation factor. Figure~\ref{fig:lattice_PRL}(d) compares this ratio with the leading massive-Dirac prediction in Eq.~\eqref{eq:PRL_Chern_ratio}, with $C\to C_M$. The full-lattice result approaches $-C_M$ at threshold and follows the Dirac scaling near threshold. At $T=0$, this interband Stokes intensity vanishes because the lower magnon band is unoccupied. For the parameters of Fig.~\ref{fig:lattice_PRL}, the absolute low-temperature signal is thermally suppressed; the cancellation above concerns the normalized response rather than its overall magnitude.

\textit{Summary and Discussion}.---
We have shown that polarization-resolved Raman spectroscopy provides an energy-resolved probe of quantum geometry in two-band magnetic systems. The counter-circular spectral sum and difference select quantum-metric and Berry-curvature contractions, respectively. Although their weights generally depend on microscopic details, $C_3$ symmetry constrains the leading Raman--velocity relation near a massive Dirac point up to a common scale. For an inversion-related Dirac pair in the $C=\pm1$ class, this scale cancels from the normalized counter-circular asymmetry, allowing the Chern number to be extracted at Raman threshold. We demonstrate this principle for Kitaev Majorana fermions and topological magnons on the honeycomb lattice, which can also be applied to various quantum magnets with Dirac quasiparticle excitations~\cite{CaiL020402, Zhu7532}.

From an experimental perspective, the Raman threshold can be estimated from existing measurements. For $\alpha$-RuCl$_3$, the field-dependent Majorana gap inferred from specific heat~\cite{Tanaka429,Imamura011045} is of order $0.9$--$1.3$ meV near $10$ T, corresponding to a Raman threshold of approximately $1.8$--$2.6$ meV; this motivates counter-circular Raman measurements over a shift range of roughly $5$--$10$ meV, already accessible in field-dependent Raman experiments~\cite{Wulferding1603}. For CrI$_3$, the magnon Dirac gap, corresponding to a Raman threshold, of approximately $2.8$--$4$ meV observed by inelastic neutron scattering~\cite{Chen041028,Chen031047} suggests targeting counter-circular Raman shifts of roughly $3$--$8$ meV. Circularly polarized magneto-Raman spectroscopy has already been demonstrated in this material~\cite{Cenker2021}.

The two counter-circular spectra $I_{RL}(\Omega)$ and $I_{LR}(\Omega)$, measured under identical conditions, yield the normalized asymmetry
\begin{align}
\nonumber
\frac{I_{RL}(\Omega)-I_{LR}(\Omega)}
     {I_{RL}(\Omega)+I_{LR}(\Omega)},
\end{align}
which can be compared directly with Eq.~\eqref{eq:PRL_Chern_ratio}. 
More generally, analogous geometric relations may arise when crystalline symmetries constrain Raman and velocity vertices up to a common scale and an orthogonal transformation. Our results thus establish a spectroscopic route to the Chern number of charge-neutral quasiparticles, for the symmetry-constrained massive Dirac class identified here, without absolute-intensity calibration or frequency integration.

\begin{acknowledgments}
We thank Dirk Wulferding and Kwang Yong Choi for fruitful discussions. This work was supported by JSPS KAKENHI, Grants No. 24H02231 (Y.Y. and T.M.), 25K23353 (Y.Y.), and 23K25816, 23K17665, 24K00568 (T.M.).
D.Y.K. and B.-J.Y. were supported by Samsung Science and Technology Foundation under Projects No. SSTF-BA2601-02, National Research Foundation of Korea (NRF) funded by the Korean government (MSIT), Grants No. RS-2021-NR060087 and No. RS-2025-00562579, Global Research Development-Center (GRDC) Cooperative Hub Program through the NRF funded by the MSIT, Grant No. RS-2023-00258359, and Global-LAMP program of the NRF funded by the Ministry of Education, Grant No. RS-2023-00301976.
\end{acknowledgments}

\bibliography{main}

\appendix


\onecolumngrid
\begin{center}
\textbf{End Matter}
\end{center}
\twocolumngrid


\setcounter{equation}{0}
\renewcommand{\theequation}{A\arabic{equation}}
\renewcommand{\theHequation}{A\arabic{equation}}

\textit{Appendix A: $C_3$ constraint on the Dirac Hamiltonian and Raman vertex.—}
Here we derive the symmetry relation used in Eq.~\eqref{eq:PRL_C3_theorem}. For the honeycomb realizations, the graphene-type origin of the Dirac points was described in the main text; here we derive the local $C_3$ constraint on the resulting Dirac doublet.

In a fixed lattice Bloch basis, the Hamiltonian satisfies
\begin{align}
D_{C_3}(\bm k)\, h(\bm k)\, D_{C_3}^{\dagger}(\bm k) = h(C_3\bm k), 
\label{eq:EM_A1}
\end{align}
where $D_{C_3}(\bm k)$ is the Bloch representation of the threefold rotation. At a valley $\bm K$ invariant under $C_3$ modulo a reciprocal lattice vector, $C_3\bm K=\bm K+\bm G_K$. Because Bloch bases at $\bm k$ and $\bm k+\bm G$ may differ by a reciprocal-lattice gluing transformation $V_{\bm G}$, $h(\bm k+\bm G) = V_{\bm G}\, h(\bm k)\,V_{\bm G}^{\dagger}$.

We construct the local $\bm k\cdot \bm p$ theory in a fixed two-state basis at $\bm K$. After reciprocal-lattice gluing, the local symmetry transformation in
a momentum-dependent lattice Bloch basis is generally itself momentum-dependent. Defining $U_K(C_3,\bm q)\equiv V_{\bm G_K}^{\dagger} D_{C_3}(\bm K+\bm q)$,
with $\bm q=\bm k-\bm K$ and $U_K(C_3,\bm 0)\equiv U_K$, the local covariance relation is
\begin{align}
U_K(C_3,\bm q)\, h_K^{(0)}(\bm q)\, U_K^{\dagger}(C_3,\bm q) = h_K^{(0)}(C_3\bm q),
\label{eq:EM_A3}
\end{align}
where $(0)$ denotes the gapless reference system.

An important point is that the momentum dependence of $U_K(C_3,\bm q)$ does not modify the linear $k\cdot p$ constraint. To see this explicitly, write
$U_K(C_3,\bm q) =U_K+q_iU_K^{(i)}+O(q^2)$ and $h_K^{(0)}(\bm q)=\epsilon_K\sigma_0+q_jh_j+O(q^2)$.
Expanding the left-hand side of Eq.~\eqref{eq:EM_A3} gives
\begin{align}
&U_K(C_3,\bm q)\,
h_K^{(0)}(\bm q)\,
U_K^{\dagger}(C_3,\bm q)-\epsilon_K\sigma_0
\nonumber\\
&=
q_jU_Kh_jU_K^\dagger
+\epsilon_K q_i
\left(
U_K^{(i)}U_K^\dagger
+
U_KU_K^{(i)\dagger}
\right)
+O(q^2)
\nonumber\\
&=
q_jU_Kh_jU_K^\dagger
+O(q^2),
\label{eq:EM_A4}
\end{align}
where we used the unitarity of $U_K(C_3,\bm q)$, which implies $U_K^{(i)}U_K^\dagger+U_KU_K^{(i)\dagger}=0$. Thus the terms generated by the $\bm q$ dependence of the symmetry transformation cancel at linear order. Crucially, this cancellation uses $h_K^{(0)}(\bm 0)=\epsilon_K\sigma_0$, i.e., the exact degeneracy at the Dirac crossing. Equation~\eqref{eq:EM_A3} therefore reduces, to linear order, to
\begin{align}
U_K h_K^{(0)}(\bm q) U_K^\dagger = h_K^{(0)}(C_3\bm q)+O(q^2).
\label{eq:EM_A5}
\end{align}

Diagonalizing the constant little-group representation $U_K$, let $\rho_1$ and $\rho_2$ denote the two $C_3$ eigenvalues of the Dirac doublet. Because the gapless doublet has a nonvanishing linear Dirac coupling, Eq.~\eqref{eq:EM_A5} requires its off-diagonal operator to transform in the same representation as an in-plane linear momentum. Since $q_\pm\equiv q_x\pm iq_y\mapsto\omega^{\pm1}q_\pm$, this requires $\rho_1/\rho_2=\omega^{\pm1}$. For the positive-chirality convention used in the main text, we choose $\rho_1/\rho_2=\omega^{-1}$. Up to an overall phase,
\begin{align}
U_K &= e^{i\phi_K} \exp\!\left(-i\frac{\pi}{3}\sigma_z\right),
&
U_K^\dagger\sigma_\pm U_K &= \omega^{\pm1}\sigma_\pm,
\label{eq:EM_A6}
\end{align}
where $\sigma_\pm\equiv(\sigma_x\pm i\sigma_y)/2$. The symmetry-allowed linear Hamiltonian of the gapless system is therefore
\begin{align}
h_K^{(0)}(\bm q) = \epsilon_K\sigma_0 +c_Kq_-\sigma_+ +c_K^*q_+\sigma_- +O(q^2).
\label{eq:EM_A7}
\end{align}
A relative phase rotation of the two basis states, which commutes with $U_K$, can be used to choose $c_K=v_K>0$.

A $C_3$-preserving gap-opening perturbation can now be classified in the same fixed basis. Because the two $C_3$ eigenvalues are distinct, a Hermitian
zeroth-order perturbation commuting with $U_K$ is a linear combination of $\sigma_0$ and $\sigma_z$. Absorbing the scalar contribution into $\epsilon_K$, we obtain the massive Dirac Hamiltonian
\begin{align}
h_K(\bm q) = \epsilon_K\sigma_0 +v_K(q_x\sigma_x+q_y\sigma_y) +m_K\sigma_z +O(q^2).
\label{eq:EM_A8}
\end{align}
All subsequent low-energy Hamiltonian and Raman vertices are expressed in this same fixed symmetry-adapted $\bm k\cdot \bm p$ basis. In this representation, any mass-dependent correction to the symmetry-allowed linear off-diagonal term can be absorbed into the coefficient $v_K$, while $C_3$ continues to forbid independent $q_i\sigma_0$ and $q_i\sigma_z$ terms.

We next turn to the Raman vertex. The essential point is that under $C_3$ the in-plane symmetric-traceless Raman doublet transforms in the same real
two-dimensional representation as the velocity doublet. At the valley center, its Cartesian components obey
\begin{align}
U_K^\dagger R_{\alpha\beta}^{(0)}U_K = (C_3)_{\alpha\alpha'}(C_3)_{\beta\beta'}R_{\alpha'\beta'}^{(0)},
\label{eq:EM_A9}
\end{align}
where the Cartesian indices refer to the same oriented physical in-plane frame used in the main text. For $R_B=R_{XX}-R_{YY}$ and $R_C=R_{XY}+R_{YX}$,
the combinations $R_B\pm iR_C$ carry angular momentum $\pm2$. Under a $2\pi/3$ rotation, $R_B+iR_C \mapsto \omega^{-1}(R_B+iR_C)$ and $R_B-iR_C
\mapsto \omega(R_B-iR_C)$.

Comparison with Eq.~\eqref{eq:EM_A6}, together with Hermiticity, gives the most general nonvanishing zeroth-order symmetric-traceless Raman doublet,
\begin{align}
R_B^{(0)}+iR_C^{(0)} &= \gamma_K\sigma_-, &
R_B^{(0)}-iR_C^{(0)} &= \gamma_K^*\sigma_+.
\label{eq:EM_A10}
\end{align}
Writing $\gamma_K=2\alpha_Ke^{i\vartheta_K}$ with $\alpha_K=|\gamma_K|/2>0$, we obtain
\begin{align}
\begin{pmatrix}
R_B^{(0)}\\
R_C^{(0)}
\end{pmatrix}
&=
\alpha_K O_K
\begin{pmatrix}
\sigma_x\\
\sigma_y
\end{pmatrix},
&
O_K
&=
\begin{pmatrix}
\cos\vartheta_K & \sin\vartheta_K\\
\sin\vartheta_K & -\cos\vartheta_K
\end{pmatrix},
\label{eq:EM_A11}
\end{align}
where $O_K^{\mathsf T}O_K=\mathbbm 1_2$ and $\det O_K=-1$. If $\gamma_K=0$, the leading Raman doublet vanishes and the normalized low-energy readout is not defined at this order.

From Eq.~\eqref{eq:EM_A8}, the leading velocity vertices are $v_x^{(0)}=v_K\sigma_x$ and $v_y^{(0)}=v_K\sigma_y$. Hence
\begin{align}
\begin{pmatrix}
R_B^{(0)}\\
R_C^{(0)}
\end{pmatrix}
=
\lambda_K O_K
\begin{pmatrix}
v_x^{(0)}\\
v_y^{(0)}
\end{pmatrix},
\qquad
\lambda_K\equiv\frac{\alpha_K}{v_K}>0 .
\label{eq:EM_A12}
\end{align}
Taking interband matrix elements gives Eq.~\eqref{eq:PRL_C3_theorem}.

The two rows of $\lambda_KO_K$ are therefore the coefficients $\lambda_B^i$ and $\lambda_C^i$ entering ${\cal G}_\Sigma$ and ${\cal F}_\Delta$ in Eq.~\eqref{eq:PRL_QGT_spectra}. Orthogonality and $\det O_K=-1$ imply $\lambda_B^i\lambda_B^j+\lambda_C^i\lambda_C^j=\lambda_K^2\delta_{ij}$ and $\lambda_B^x\lambda_C^y-\lambda_B^y\lambda_C^x =-\lambda_K^2$. Consequently,
\begin{align}
{\cal G}_\Sigma^{ij, K} = \lambda_K^2 \delta_{ij}, \qquad {\cal F}_\Delta^K = -\lambda_K^2.
\label{eq:EM_A13}
\end{align}
These are the relations quoted below Eq.~\eqref{eq:PRL_C3_theorem}. At the opposite-chirality valley $\bm K'$, the local $C_3$ representation may be chosen as the complex-conjugate one. When the velocity components are defined with respect to the same oriented Cartesian frame, repeating the construction above again
yields an orthogonal Raman--velocity transformation with $\det O_{K'}=-1$. The relative minus sign between the Raman-selected Berry-curvature contraction and $F_{xy}$ is therefore the same at the two valleys.


\setcounter{equation}{0}
\renewcommand{\theequation}{B\arabic{equation}}
\renewcommand{\theHequation}{B\arabic{equation}}

\textit{Appendix B: Massive-Dirac Raman spectra and the Chern-number relation.—}
We now derive Eq.~\eqref{eq:PRL_Chern_ratio} from the $C_3$-constrained massive-Dirac Raman structure obtained in Appendix A. Consider first a single positive-chirality massive Dirac cone. Dropping the scalar energy shift, which affects neither the eigenstates nor the interband separation, we write
\begin{align}
h_K(\bm q) &= v_K(q_x\sigma_x+q_y\sigma_y)+m_K\sigma_z,
\nonumber\\
{\rm Tr}\,g^K(\bm q) &= \frac{v_K^2(E_K^2+m_K^2)}{4E_K^4},
\qquad
F_{xy}^K(\bm q) = \frac{m_Kv_K^2}{2E_K^3},
\label{eq:EM_B1}
\end{align}
where $E_K(\bm q)=\sqrt{v_K^2q^2+m_K^2}$ and the interband separation is $\Delta_K(\bm q)=2E_K(\bm q)$.

Using the relations derived in Appendix A, the counter-circular sum and difference for this cone are
\begin{align}
\frac{I_\Sigma^K(\Omega)}{\Omega^2} &= 2{\cal N}_R\lambda_K^2
\int_{\bm q}
{\rm Tr}\,g^K(\bm q)
\delta[\Omega-2E_K(\bm q)],
\nonumber\\
\frac{I_\Delta^K(\Omega)}{\Omega^2}
&=
-2{\cal N}_R\lambda_K^2
\int_{\bm q}
F_{xy}^K(\bm q)
\delta[\Omega-2E_K(\bm q)],
\label{eq:EM_B2}
\end{align}
where $\int_{\bm q}\equiv\int d^2q/(2\pi)^2$. With $\Omega_{\rm th}^{K}=2|m_K|$, direct radial integration of Eq.~\eqref{eq:EM_B2} gives
\begin{align}
I_\Sigma^K(\Omega)
&=
\frac{{\cal N}_R\lambda_K^2}{4\pi}
\frac{\Omega^2+(\Omega_{\rm th}^K)^2}{\Omega}
\Theta(\Omega-\Omega_{\rm th}^K),
\nonumber\\
I_\Delta^K(\Omega)
&=
-\frac{{\cal N}_R\lambda_K^2}{2\pi}
\operatorname{sgn}(m_K)\,
\Omega_{\rm th}^K
\Theta(\Omega-\Omega_{\rm th}^K).
\label{eq:EM_B3}
\end{align}
Their ratio is therefore
\begin{align}
\frac{I_\Delta^K(\Omega)}
{I_\Sigma^K(\Omega)}
=
-\operatorname{sgn}(m_K)
\frac{2\Omega_{\rm th}^K\Omega}
{\Omega^2+(\Omega_{\rm th}^K)^2}.
\label{eq:EM_B4}
\end{align}
The common Raman normalization ${\cal N}_R$, Raman scale $\lambda_K$, and Dirac velocity $v_K$ have canceled.

We next combine two valleys exchanged by inversion. All momentum components are expressed in the same oriented physical Cartesian frame. Let $-\bm K=\bm K'+\bm G_P$, where $\bm G_P$ is a reciprocal lattice vector. Denoting the Bloch representation of inversion by $D_P(\bm k)$, we define the locally glued inversion sewing matrix $U_P(\bm q) \equiv V_{\bm G_P}^{\dagger}D_P(\bm K+\bm q)$. It relates the local two-band bases at $\bm K+\bm q$ and $\bm K'-\bm q$.
Inversion covariance gives
\begin{align}
U_P(\bm q)\, h_K(\bm q)\, U_P^\dagger(\bm q) =h_{K'}(-\bm q).
\label{eq:EM_B5}
\end{align}
The two valleys therefore have identical eigenvalue dispersions. Comparing the leading Dirac spectra,
\begin{align}
\Delta_\tau(\bm q) = 2\sqrt{v_\tau^2q^2+m_\tau^2}, \qquad \tau=K,K',
\label{eq:EM_B6}
\end{align}
gives $v_K=v_{K'}\equiv v$ and $|m_K|=|m_{K'}|\equiv m$. Importantly, inversion fixes the magnitude of the Dirac mass, but not the sign of the coefficient $m_\tau$ defined in independently chosen valley bases.

The Raman tensor is likewise inversion covariant. Because its two polarization indices transform as polar-vector indices, the two inversion signs cancel:
\begin{align}
U_P(\bm q)\, R_{\alpha\beta}^K(\bm q)\, U_P^\dagger(\bm q) = R_{\alpha\beta}^{K'}(-\bm q).
\label{eq:EM_B7}
\end{align}
At the valley centers, Eq.~\eqref{eq:EM_B7} implies $|\gamma_K|=|\gamma_{K'}|$, and hence $\alpha_K=\alpha_{K'}$. Since $\lambda_\tau=\alpha_\tau/v_\tau>0$, together with Eq.~\eqref{eq:EM_B6} this gives $\lambda_K=\lambda_{K'}\equiv\lambda$. Thus, the two valleys share the common Raman scale and threshold, with
$|m_K|=|m_{K'}|=m$, $\lambda_K=\lambda_{K'}=\lambda$, and $\Omega_{\rm th}=2m$.

For the lower band, inversion symmetry gives
\begin{align}
{\rm Tr}\,g^{K'}(\bm q) &= {\rm Tr}\,g^K(-\bm q),
&
F_{xy}^{K'}(\bm q) &= F_{xy}^{K}(-\bm q).
\label{eq:EM_B8}
\end{align}
The Berry curvature is therefore inversion even in the common oriented physical frame. Together with the equal Raman scales and the common orientation-reversing Raman--velocity relation established in Appendix A, Eq.~\eqref{eq:EM_B8} implies that the two valley contributions to both $I_\Sigma$ and $I_\Delta$ add.

For the positive-chirality convention adopted at $\bm K$, the continuum Dirac cone carries $\frac{1}{2\pi}\int d^2q\,F_{xy}^K(\bm q) =
\frac{1}{2}\operatorname{sgn}(m_K)$. This half-integer result is the weak-mass contribution associated with a Dirac valley. For a gapped lattice system adiabatically connected, without additional gap closings, to a chiral-symmetric parent whose only touchings are the inversion-related linear Dirac pair $K$ and $K'$, the weak-mass limit localizes the Berry-curvature contribution at these valleys. Since inversion makes the two valley contributions equal, the lower-band Chern number $C=(2\pi)^{-1}\int_{\rm BZ}d^2k\,F_{xy}(\bm k)$ is fixed throughout the adiabatically connected gapped phase as
\begin{align}
C=\operatorname{sgn}(m_K)=\pm1,
\qquad
I_{\Sigma,\Delta}(\Omega)
=
2I_{\Sigma,\Delta}^K(\Omega),
\label{eq:EM_B10}
\end{align}
where the second relation holds within the frequency window in which the Raman shell is confined to the leading Dirac neighborhoods. Combining Eqs.~\eqref{eq:EM_B4} and~\eqref{eq:EM_B10} gives Eq.~\eqref{eq:PRL_Chern_ratio}.


\clearpage
\onecolumngrid

\begin{center}

{\large\bfseries
Supplemental Material for\\[2mm]
``Probing the Dirac Quantum Geometry in Quantum Magnets
from Polarized Raman Spectroscopy''
\par}

\vspace{4mm}

Yuki Yamazaki$^{1}$,
Doh Young Kim$^{2,3,4}$,
Takahiro Morimoto$^{1,*}$,
and Bohm-Jung Yang$^{2,3,4,\dagger}$

\vspace{3mm}

{\it \small
$^{1}$Department of Physics, Graduate School of Science,
Kyoto University, Kyoto 606-8502, Japan

$^{2}$Department of Physics and Astronomy,
Seoul National University, Seoul 08826, Korea

$^{3}$Center for Theoretical Physics (CTP),
Seoul National University, Seoul 08826, Korea

$^{4}$Institute of Applied Physics,
Seoul National University, Seoul 08826, Korea
}

\vspace{2mm}

{\small
$^{*}$morimoto@scphys.kyoto-u.ac.jp
\qquad
$^{\dagger}$bjyang@snu.ac.kr
}

\end{center}

\vspace{4mm}

\setcounter{section}{0}
\setcounter{subsection}{0}
\setcounter{equation}{0}
\setcounter{figure}{0}
\setcounter{table}{0}

\renewcommand{\thesection}{S\arabic{section}}
\renewcommand{\thesubsection}{\Alph{subsection}}
\renewcommand{\theequation}{S\arabic{section}.\arabic{equation}}
\renewcommand{\thefigure}{S\arabic{section}.\arabic{figure}}
\renewcommand{\thetable}{S\arabic{section}.\arabic{table}}

\renewcommand{\theHsection}{supp.section.\arabic{section}}
\renewcommand{\theHsubsection}
  {supp.subsection.\arabic{section}.\arabic{subsection}}
\renewcommand{\theHequation}
  {supp.equation.\arabic{section}.\arabic{equation}}
\renewcommand{\theHfigure}
  {supp.figure.\arabic{section}.\arabic{figure}}
\renewcommand{\theHtable}
  {supp.table.\arabic{section}.\arabic{table}}

\newcommand{\suppsection}[1]{%
  \refstepcounter{section}%
  \section*{\thesection.\quad #1}%
}

\newcommand{\suppsubsection}[1]{%
  \refstepcounter{subsection}%
  \subsection*{\thesubsection.\quad #1}%
}

\suppsection{Magnetic Raman vertices}
\label{sec:SM_Raman_vertices}

In this section, we summarize the magnetic Raman vertices used in the Kitaev-Majorana and topological-magnon calculations below. We consider two complementary constructions: the conventional Loudon--Fleury (LF) exchange-scattering vertex and an effective second-order light--matter-coupling (LMC) vertex. For the class of processes considered in this work, both constructions give Hermitian Cartesian Raman components and a symmetric Raman tensor. The LF and LMC constructions need not be equivalent in general; below they are applied to the microscopic terms for which each construction is appropriate.

\suppsubsection{Loudon--Fleury Raman vertex}
\label{sec:SM_LF_vertex}

Consider a spin Hamiltonian written as a sum of exchange terms,
\begin{align}
\hat H = \sum_{(ij)}\hat H_{ij},
\label{eq:SM_spin_bond_H}
\end{align}
where $(ij)$ labels exchange-coupled bonds. The LF Raman operator is
\begin{align}
\hat R_{\rm LF} = \sum_{(ij)} (\bm e_{\rm in}\cdot\bm d_{ij}) (\bm e_{\rm out}^{*}\cdot\bm d_{ij}) \hat H_{ij},
\label{eq:SM_LF_operator}
\end{align}
up to an overall Raman-coupling constant, which is absorbed into the spectral normalization throughout this work~\cite{Fleury514,Shastry1068}. Here $\bm e_{\rm in}$ and $\bm e_{\rm out}$ are the incoming and outgoing photon polarizations and $\bm d_{ij}$ is the bond vector.

Equivalently,
\begin{align}
\hat R_{\rm LF} = e_{{\rm out},\alpha}^{*}\hat R_{\alpha\beta}^{\rm LF}e_{{\rm in},\beta},\qquad \hat R_{\alpha\beta}^{\rm LF} = \sum_{(ij)} d_{ij}^{\alpha}d_{ij}^{\beta} \hat H_{ij},
\label{eq:SM_LF_tensor}
\end{align}
where repeated in-plane Cartesian indices $\alpha,\beta=X,Y$ are summed. Since $d_{ij}^{\alpha}d_{ij}^{\beta} =d_{ij}^{\beta}d_{ij}^{\alpha}$,
\begin{align}
\hat R_{\alpha\beta}^{\rm LF} = \hat R_{\beta\alpha}^{\rm LF}.
\label{eq:SM_LF_symmetric}
\end{align}

After projection to a quadratic quasiparticle description, Eqs.~\eqref{eq:SM_spin_bond_H} and \eqref{eq:SM_LF_tensor} schematically take the following forms, up to the conventional normalization factors specified below:
\begin{align}
\hat{H} &\longrightarrow \sum_{\bm k}\bm\Psi_{\bm k}^\dagger h(\bm k)\bm\Psi_{\bm k},
\label{eq:Hspin_to_h}
\\
\hat R_{\alpha\beta}^{\rm LF} &\longrightarrow \sum_{\bm k}\bm\Psi_{\bm k}^\dagger R_{\alpha\beta}^{\rm LF}(\bm k)\bm\Psi_{\bm k},
\label{eq:RLF_to_Rk}
\end{align}
where $\bm\Psi_{\bm k}$ denotes the corresponding Majorana or magnon spinor. Thus, the effective quasiparticle Hamiltonian $h(\bm k)$ and Raman vertex $R_{\alpha\beta}^{\rm LF}(\bm k)$ act in the same effective quasiparticle Hilbert space.

\suppsubsection{Effective light--matter-coupling Raman vertex}
\label{sec:SM_LMC_vertex}

A complementary construction starts directly from an effective quasiparticle Hamiltonian $h(\bm k)$. When the effective coupling to light admits a momentum-shift representation, the second-order LMC vertex is obtained from second momentum derivatives of $h(\bm k)$~\cite{Bostrom026701,Koller194425,Liu2026}. Schematically,
\begin{align}
\frac{1}{2}
\left[h(\bm k-\bm A) + h(\bm k+\bm A) \right] = h(\bm k) + \frac{1}{2} A_\alpha A_\beta \partial_{k_\alpha}\partial_{k_\beta} h(\bm k) + O(A^4),
\label{eq:SM_LMC_expansion}
\end{align}
so that, up to an overall coupling constant,
\begin{align}
R_{\alpha\beta}^{\rm LMC}(\bm k) \propto \partial_{k_\alpha}\partial_{k_\beta} h(\bm k).
\label{eq:SM_LMC_tensor}
\end{align}
This momentum-shift form should be understood as an effective representation of the magnetic light--matter coupling rather than as a literal electromagnetic minimal coupling of charge-neutral quasiparticles.

All momentum derivatives in Eq.~\eqref{eq:SM_LMC_tensor} are taken in the same fixed Bloch representation, including the same orbital or sublattice embedding used to define the effective quasiparticle Hamiltonian. For a smooth Hermitian $h(\bm k)$, the Cartesian LMC components are Hermitian and satisfy
\begin{align}
R_{\alpha\beta}^{\rm LMC}(\bm k) = R_{\beta\alpha}^{\rm LMC}(\bm k).
\label{eq:SM_LMC_symmetric}
\end{align}

The LF and LMC constructions are not assumed to be generally equivalent. Under appropriate microscopic conditions they can represent the same underlying virtual-hopping process and yield proportional effective vertices~\cite{Liu2026}. In the applications below, we use the LF construction for the nearest-neighbor exchange contribution. For the field-induced three-spin term of the Kitaev-Majorana model and the Dzyaloshinskii--Moriya contribution of the magnon model, we use the effective LMC construction.

For later use, we introduce the symmetric-traceless Raman components
\begin{align}
R_B &\equiv R_{XX}-R_{YY}, &
R_C &\equiv R_{XY}+R_{YX}.
\label{eq:SM_BC_def}
\end{align}
With
\begin{align}
\bm e_R = \frac{\hat{\bm X}+i\hat{\bm Y}}{\sqrt2}, \qquad \bm e_L = \frac{\hat{\bm X}-i\hat{\bm Y}}{\sqrt2},
\label{eq:SM_circular_vectors}
\end{align}
and with the first polarization label denoting the incoming photon and the second the outgoing photon, the symmetric Raman tensor gives
\begin{align}
2R_{RL} &= R_B+iR_C, &
2R_{LR} &= R_B-iR_C.
\label{eq:SM_counter_circular_vertices}
\end{align}
The common factor $1/2$ can be absorbed into the overall Raman normalization.
\setcounter{equation}{0}

\suppsection{Kitaev Majorana model}
\label{sec:SM_Kitaev}

In this section, we provide the lattice conventions, the flux-free Majorana Bloch Hamiltonian, and the Raman vertices used in the main text. We then derive explicitly the relation between the isotropized local Dirac coordinates and the physical Cartesian momentum frame and show how the microscopic Raman vertices realize the $C_3$-constrained Raman--velocity relation. We also clarify how the physical two-Majorana creation process is represented by the negative-to-positive-energy interband matrix element of the Raman vertex. Finally, we examine the dependence of the full-lattice Raman response on the microscopic coupling $\zeta_R$ and demonstrate the robustness of the normalized threshold asymmetry.

\suppsubsection{Majorana representation and effective Hamiltonian}
\label{sec:SM_Kitaev_H}

We first derive the two-band Majorana Hamiltonian used in the main text from the spin Hamiltonian. We consider the isotropic Kitaev-Majorana model~\cite{Kitaev2, Motome012002}
\begin{align}
\hat H_K = -J\sum_{\mu=x,y,z}\sum_{\langle ij\rangle_\mu}S_i^\mu S_j^\mu ,
\label{eq:SM_Kitaev_spin_H}
\end{align}
together with the field-induced three-spin interaction
\begin{align}
\hat H_\eta = 2\eta \sum_{\langle ijk\rangle_{\alpha\beta}} S_i^\alpha S_j^\gamma S_k^\beta ,
\label{eq:SM_Kitaev_three_spin}
\end{align}
where $\langle ijk\rangle_{\alpha\beta}$ denotes an oriented two-bond path and $(\alpha,\beta,\gamma)$ is a permutation of $(x,y,z)$. The coupling $\eta$ is proportional to $h_xh_yh_z$ to leading order in a weak magnetic field. 

\begin{figure}[t]
\centering
\includegraphics[width=0.3\linewidth]{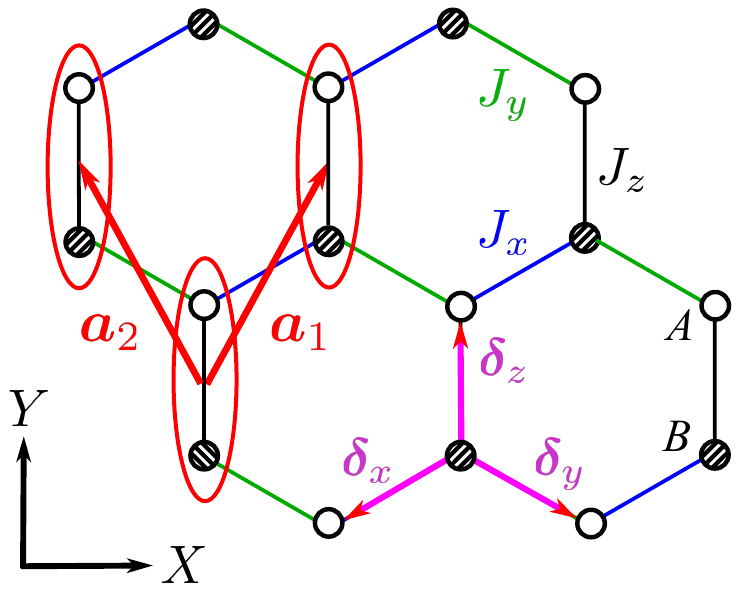}
\caption{Honeycomb-lattice convention for the Kitaev-Majorana model. The nearest-neighbor $x$, $y$, and $z$ bonds, the two sublattices $A$ and $B$, the primitive lattice vectors $\bm a_1$ and $\bm a_2$, the nearest-neighbor vectors $\bm\delta_x,\bm\delta_y,\bm\delta_z$, and the physical Cartesian axes $(X,Y)$ are shown.}
\label{fig:SM_Kitaev_lattice}
\end{figure}

The lattice convention is shown in Fig.~\ref{fig:SM_Kitaev_lattice}. We use
\begin{align}
\bm a_1 = \left(\frac12,\frac{\sqrt3}{2}\right),
\qquad
\bm a_2 = \left(-\frac12,\frac{\sqrt3}{2}\right),
\label{eq:SM_Kitaev_lattice_vectors}
\end{align}
defined in the physical Cartesian frame $(X,Y)$, and introduce
\begin{align}
p_i \equiv \bm k\cdot\bm a_i, \qquad i=1,2.
\label{eq:SM_Kitaev_pi}
\end{align}

To obtain the Majorana representation, we regard the honeycomb lattice as a brick-wall lattice composed of one-dimensional chains of alternating $x$ and $y$ bonds and choose a Jordan--Wigner ordering along these chains. Introducing spinless fermions $a_j$, the spin-$1/2$ operators are written as
\begin{align}
S_j^+ &= a_j^\dagger \prod_{\ell<j} (1-2n_\ell), \qquad
S_j^-= \prod_{\ell<j} (1-2n_\ell)\, a_j,
\nonumber\\
S_j^z &= n_j-\frac12, \qquad n_j=a_j^\dagger a_j.
\label{eq:SM_Kitaev_JW}
\end{align}
Here $\ell<j$ refers to the chosen one-dimensional ordering.

For a unit cell at
\begin{align}
\bm r = r_1\bm a_1+r_2\bm a_2, \qquad r_1,r_2\in\mathbb Z,
\end{align}
we decompose the spinless fermions on the two sublattices into Majorana operators using the sublattice-dependent convention
\begin{align}
\gamma_{\bm r,A} &= a_{\bm r,A} + a_{\bm r,A}^\dagger, & \bar\gamma_{\bm r,A} &= -i\left(a_{\bm r,A} - a_{\bm r,A}^\dagger \right), \nonumber\\
\gamma_{\bm r,B} &= -i\left(a_{\bm r,B} -a_{\bm r,B}^\dagger\right), & \bar\gamma_{\bm r,B} &= a_{\bm r,B} + a_{\bm r,B}^\dagger.
\label{eq:SM_Kitaev_JW_Majorana}
\end{align}
They satisfy
\begin{align}
\gamma_{\bm r,s}^\dagger = \gamma_{\bm r,s}, \quad \bar\gamma_{\bm r,s}^\dagger = \bar\gamma_{\bm r,s},
\quad \{\gamma_{\bm r,s}, \gamma_{\bm r',s'} \}= 2\delta_{\bm r\bm r'}\delta_{ss'},
\label{eq:SM_Kitaev_Majorana_algebra}
\end{align}
and analogously for $\bar\gamma$.

With the bond convention of Fig.~\ref{fig:SM_Kitaev_lattice}, the Kitaev Hamiltonian becomes
\begin{align}
\hat H_K = \frac{iJ}{4} \sum_{\bm r} \Big[&\gamma_{\bm r,B}\gamma_{\bm r-\bm a_1,A}+\gamma_{\bm r,B}\gamma_{\bm r-\bm a_2,A}+u_{\bm r} \gamma_{\bm r,B}\gamma_{\bm r,A}\Big],
\label{eq:SM_Kitaev_realspace_Majorana}
\end{align}
where
\begin{align}
u_{\bm r} = i\bar\gamma_{\bm r,A}\bar\gamma_{\bm r,B}, \qquad u_{\bm r}^2=1.
\label{eq:SM_Kitaev_u}
\end{align}
The $u_{\bm r}$ commute with one another and with $\hat H_K$,
\begin{align}
[u_{\bm r},u_{\bm r'}] = [u_{\bm r},\hat H_K] = 0,
\end{align}
and therefore act as static $\mathbb Z_2$ gauge variables on the $z$ bonds. Their products determine the plaquette fluxes. For the ground-state flux sector we choose the translationally invariant representative
\begin{align}
u_{\bm r}=+1.
\label{eq:SM_Kitaev_fluxfree}
\end{align}

The field-induced three-spin interaction Eq.~\eqref{eq:SM_Kitaev_three_spin} preserves the flux sector. In the Majorana representation it produces oriented next-nearest-neighbor hopping of the itinerant Majoranas $\gamma_{\bm r,s}$. Thus, in the flux-free sector, $\hat H_K+\hat H_\eta$ is a quadratic Hamiltonian entirely in terms of the itinerant Majoranas.

We Fourier transform according to
\begin{align}
\gamma_{\bm r,s} = \sqrt{\frac{2}{N}} \sum_{\bm k} e^{i\bm k\cdot\bm r} \gamma_{\bm k,s}, \qquad s=A,B.
\label{eq:SM_Kitaev_Fourier}
\end{align}
Here $N$ denotes the number of unit cells, and the momentum sum runs over the full Brillouin zone. The Majorana condition gives
\begin{align}
\gamma_{\bm k,s}^{\dagger} = \gamma_{-\bm k,s}.
\label{eq:SM_Kitaev_reality}
\end{align}
The three nearest-neighbor Majorana hoppings generate the structure
factor
\begin{align}
f(\bm k) = 1+e^{-ip_1}+e^{-ip_2}.
\label{eq:SM_Kitaev_structure_factor}
\end{align}
For later convenience, we perform a momentum-independent basis transformation $\gamma_{\bm r,B}\to-\gamma_{\bm r,B}$, leaving $\gamma_{\bm r,A}$ unchanged. In this basis, the flux-free Hamiltonian takes the form
\begin{align}
\hat H_{\rm Maj} = \frac12 \sum_{\bm k} \bm\gamma_{\bm k}^{\dagger} h_{\rm Maj}(\bm k) \bm\gamma_{\bm k}, \qquad \bm\gamma_{\bm k}
=
\begin{pmatrix}
\gamma_{\bm k,A}\\
-\gamma_{\bm k,B}
\end{pmatrix},
\label{eq:SM_Kitaev_Bloch_H}
\end{align}
with
\begin{align}
h_{\rm Maj}(\bm k) = d_x(\bm k)\sigma_x + d_y(\bm k)\sigma_y + d_z(\bm k)\sigma_z.
\label{eq:SM_Kitaev_h}
\end{align}

The nearest-neighbor term gives
\begin{align}
d_x(\bm k)+id_y(\bm k) = -i\frac{J}{2}\left(1+e^{-ip_1}+e^{-ip_2}\right).
\label{eq:SM_Kitaev_dxy}
\end{align}

The oriented next-nearest-neighbor hopping generated by $\hat H_\eta$ contributes with opposite signs on the two sublattices. For the orientation convention used here, its Fourier transform is~\cite{Kitaev2, Motome012002}
\begin{align}
d_z(\bm k) = \eta \left[-\sin p_1+\sin p_2+\sin(p_1-p_2)\right].
\label{eq:SM_Kitaev_dz}
\end{align}
Equations~\eqref{eq:SM_Kitaev_dxy} and \eqref{eq:SM_Kitaev_dz} are the effective two-band Hamiltonian used in the full-lattice calculation.

The Majorana reality condition is equivalently encoded in
\begin{align}
h_{\rm Maj}(-\bm k) = -h_{\rm Maj}^{T}(\bm k).
\label{eq:SM_Kitaev_PH}
\end{align}
The two eigenvalues of $h_{\rm Maj}(\bm k)$ are
\begin{align}
E_\pm(\bm k) = \pm E(\bm k), \qquad E(\bm k) = |\bm d(\bm k)|,
\label{eq:SM_Kitaev_eigenvalues}
\end{align}
and the negative-to-positive-energy separation is
\begin{align}
\Delta(\bm k) = E_+(\bm k)-E_-(\bm k) = 2E(\bm k).
\label{eq:SM_Kitaev_excitation}
\end{align}

\suppsubsection{Loudon--Fleury and field-induced Raman vertices}
\label{sec:SM_Kitaev_Raman}

The nearest-neighbor magnetic Raman response is described by the Loudon--Fleury operator
\begin{align}
\hat R_{\rm LF} = -J\sum_{\mu=x,y,z} \sum_{\langle ij\rangle_\mu} (\bm e_{\rm in}\cdot\bm\delta_\mu) (\bm e_{\rm out}^{*}\cdot\bm\delta_\mu) S_i^\mu S_j^\mu .
\label{eq:SM_Kitaev_LF}
\end{align}
We choose the three nearest-neighbor bond vectors as
\begin{align}
\bm\delta_x &= \left(-\frac12,-\frac{1}{2\sqrt3}\right), & \bm\delta_y &= \left( \frac12,-\frac{1}{2\sqrt3} \right), &
\bm\delta_z &= \left( 0,\frac{1}{\sqrt3} \right).
\label{eq:SM_Kitaev_bond_vectors}
\end{align}

Projection of Eq.~\eqref{eq:SM_Kitaev_LF} to the flux-free Majorana sector gives
\begin{align}
R_{{\rm LF},P}^{\rm Maj}(\bm k) =
\begin{pmatrix}
0
&
z_{{\rm LF},P}^{*}(\bm k)
\\
z_{{\rm LF},P}(\bm k)
&
0
\end{pmatrix},
\qquad
P=B,C,
\label{eq:SM_Kitaev_RLF_matrix}
\end{align}
with
\begin{align}
z_{{\rm LF},B}(\bm k) &= -\frac{iJ}{12} \left(e^{-ip_1} + e^{-ip_2} -2 \right),
\label{eq:SM_Kitaev_RLF_B}
\\
z_{{\rm LF},C}(\bm k) &= -\frac{iJ}{4\sqrt3} \left( e^{-ip_1} - e^{-ip_2} \right).
\label{eq:SM_Kitaev_RLF_C}
\end{align}
Here we use the same basis convention as in Eq.~\eqref{eq:SM_Kitaev_Bloch_H}.

Following the effective LMC construction for the field-induced three-spin term~\cite{Koller194425}, we define
\begin{align}
R_{\eta,\alpha\beta}^{\rm Maj}(\bm k) = - \partial_{k_\alpha} \partial_{k_\beta} d_z(\bm k) \,\sigma_z, \qquad \alpha,\beta=X,Y.
\label{eq:SM_Kitaev_Reta_definition}
\end{align}
The LF and LMC vertices are related by $R_{\mathrm{LF},\alpha\beta}=-R_{\mathrm{LMC},\alpha\beta}$ when the momentum derivatives are taken in a Bloch basis incorporating the physical site positions~\cite{Koller194425}. In the unit-cell convention used here, the corresponding derivatives of the nearest-neighbor Hamiltonian require the sublattice-embedding correction. By contrast, the field-induced three-spin term is diagonal in the sublattice basis and remains unchanged under this basis transformation. Therefore, its LMC vertex can be obtained directly from the ordinary momentum derivatives of $d_z(\boldsymbol k)\sigma_z$, as in Eq.~\eqref{eq:SM_Kitaev_Reta_definition}.
Its symmetric-traceless components are
\begin{align}
R_{\eta,B}^{\rm Maj}(\bm k) &= \eta \left[
\frac12\sin p_1 - \frac12\sin p_2 + \sin(p_1-p_2) \right] \sigma_z,
\label{eq:SM_Kitaev_Reta_B}
\\
R_{\eta,C}^{\rm Maj}(\bm k) &= -\frac{\sqrt3\eta}{2} \left( \sin p_1+\sin p_2 \right) \sigma_z.
\label{eq:SM_Kitaev_Reta_C}
\end{align}

The full Raman vertices used in the numerical calculation are
\begin{align}
\hat R_P &= \frac12 \sum_{\bm k} \bm\gamma_{\bm k}^{\dagger} R_P^{\rm Maj}(\bm k) \bm\gamma_{\bm k},
\label{eq:SM_Kitaev_Raman_operator_band} \\
R_P^{\rm Maj}(\bm k) &= R_{{\rm LF},P}^{\rm Maj}(\bm k) + \zeta_R R_{\eta,P}^{\rm Maj}(\bm k), \qquad P=B,C.
\label{eq:SM_Kitaev_full_Raman}
\end{align}
The dimensionless parameter $\zeta_R$ parametrizes the microscopic strength of the field-induced light--matter coupling relative to the nearest-neighbor LF contribution.

\suppsubsection{Dirac expansion and relation between local and physical
momentum frames}
\label{sec:SM_Kitaev_Dirac}

We next derive explicitly the low-energy expressions used in the main text. The two inequivalent valleys are
\begin{align}
\bm K:
\
(p_1,p_2)
&=
\left(
\frac{2\pi}{3},
-\frac{2\pi}{3}
\right),
\qquad
\bm K':
\
(p_1,p_2)
=
\left(
-\frac{2\pi}{3},
\frac{2\pi}{3}
\right).
\label{eq:SM_Kitaev_valleys}
\end{align}
We introduce
\begin{align}
\chi_K=+1,
\qquad
\chi_{K'}=-1,
\end{align}
so that both valleys can be written as
\begin{align}
p_1^\tau
=
\chi_\tau\frac{2\pi}{3},
\qquad
p_2^\tau
=
-\chi_\tau\frac{2\pi}{3},
\label{eq:SM_Kitaev_valley_compact}
\end{align}
where $\tau=K,K'$. For
\begin{align}
p_i=p_i^\tau+\delta p_i,
\end{align}
we define the local momentum coordinates
\begin{align}
\kappa_x &= \frac{\delta p_1+\delta p_2}{2}, \qquad
\kappa_y = \frac{\sqrt3}{2}\left(\delta p_1-\delta p_2\right).
\label{eq:SM_Kitaev_kappa}
\end{align}

It is important to distinguish these isotropized local coordinates from the physical Cartesian momentum coordinates $(k_X,k_Y)$ used to define the photon polarizations and the velocity vertices in the main text. Since $p_i=\bm k\cdot\bm a_i$,
\begin{align}
\delta p_1 = \frac12\delta k_X + \frac{\sqrt3}{2}\delta k_Y,
\qquad
\delta p_2 = -\frac12\delta k_X + \frac{\sqrt3}{2}\delta k_Y.
\label{eq:SM_Kitaev_dp_XY}
\end{align}
Combining Eqs.~\eqref{eq:SM_Kitaev_kappa} and \eqref{eq:SM_Kitaev_dp_XY} gives
\begin{align}
\begin{pmatrix}
\kappa_x\\
\kappa_y
\end{pmatrix}
=
\frac{\sqrt3}{2}
\begin{pmatrix}
0&1\\
1&0
\end{pmatrix}
\begin{pmatrix}
\delta k_X\\
\delta k_Y
\end{pmatrix}.
\label{eq:SM_Kitaev_kappa_XY}
\end{align}
Thus, the local coordinates are isotropic but have the opposite orientation to the physical Cartesian momentum frame:
\begin{align}
\det \frac{\partial(\kappa_x,\kappa_y)}{\partial(k_X,k_Y)} = -\frac34.
\label{eq:SM_Kitaev_orientation}
\end{align}

Expanding Eq.~\eqref{eq:SM_Kitaev_dxy} to linear order in $\bm\kappa$ gives
\begin{align}
d_x+i d_y
=
\frac{J}{2}\kappa_x
+
i\chi_\tau\frac{J}{2}\kappa_y
+
O(\kappa^2).
\label{eq:SM_Kitaev_dxy_expansion}
\end{align}
For the mass function,
\begin{align}
-\sin p_1+\sin p_2+\sin(p_1-p_2)
=
-\chi_\tau\frac{3\sqrt3}{2}
+
O(\kappa^2),
\label{eq:SM_Kitaev_mass_expansion}
\end{align}
where the linear term vanishes.
The local Dirac Hamiltonian therefore becomes
\begin{align}
h_\tau(\bm\kappa)
=
\frac{J}{2}
\kappa_x\sigma_x
+
\frac{J}{2}
\chi_\tau
\kappa_y\sigma_y
+
m_\tau\sigma_z
+
O(\kappa^2),
\label{eq:SM_Kitaev_Dirac}
\end{align}
with
\begin{align}
m_\tau
=
-\chi_\tau
\frac{3\sqrt3}{2}\eta.
\label{eq:SM_Kitaev_mass}
\end{align}
The two valleys thus have the same mass magnitude,
\begin{align}
|m_\tau|
=
\frac{3\sqrt3}{2}|\eta|,
\end{align}
and the negative-to-positive-energy threshold is
\begin{align}
\Omega_{\rm th}
=
2|m_\tau|
=
3\sqrt3|\eta|.
\label{eq:SM_Kitaev_threshold}
\end{align}

We next expand the microscopic Raman vertices in the same local coordinates. Equations~\eqref{eq:SM_Kitaev_RLF_B} and \eqref{eq:SM_Kitaev_RLF_C} give at the valley centers
\begin{align}
z_{{\rm LF},B}(\bm K_\tau) = \frac{iJ}{4}, \qquad
z_{{\rm LF},C}(\bm K_\tau) = -\chi_\tau\frac{J}{4},
\end{align}
and therefore
\begin{align}
R_{{\rm LF},B,\tau}^{(0)}
&=
\frac{J}{4}\sigma_y,
\label{eq:SM_Kitaev_RB0}
\\
R_{{\rm LF},C,\tau}^{(0)}
&=
-\chi_\tau\frac{J}{4}\sigma_x.
\label{eq:SM_Kitaev_RC0}
\end{align}
Using Eq.~\eqref{eq:SM_Kitaev_Dirac}, these equivalently are written as
\begin{align}
R_{{\rm LF},B,\tau}^{(0)} &= \frac{\chi_\tau}{2} \partial_{\kappa_y}h_\tau,
\\
R_{{\rm LF},C,\tau}^{(0)} &= -\frac{\chi_\tau}{2} \partial_{\kappa_x}h_\tau.
\label{eq:SM_Kitaev_R_kappa_velocity}
\end{align}

The field-induced LMC vertices vanish exactly at both valley centers,
\begin{align}
R_{\eta,B}^{\rm Maj}(\bm K_\tau) = R_{\eta,C}^{\rm Maj}(\bm K_\tau) = 0.
\label{eq:SM_Kitaev_Reta_valley_zero}
\end{align}
Their leading nonzero terms therefore are linear in the local momentum,
\begin{align}
R_{\eta,B}^{\rm Maj}(\bm K_\tau+\bm\kappa) &= -\frac{\sqrt3\eta}{2} \kappa_y\sigma_z + O(\kappa^2),
\\
R_{\eta,C}^{\rm Maj}(\bm K_\tau+\bm\kappa) &= \frac{\sqrt3\eta}{2} \kappa_x\sigma_z + O(\kappa^2).
\label{eq:SM_Kitaev_Reta_expansion}
\end{align}

To compare the microscopic Raman vertices with the general Raman--velocity relation in the main text, we transform the velocity vertices back to the common oriented physical Cartesian frame. Equation~\eqref{eq:SM_Kitaev_kappa_XY} gives
\begin{align}
v_X^\tau
\equiv
\partial_{k_X}h_\tau
&=
\frac{\sqrt3}{2}
\partial_{\kappa_y}h_\tau
=
\chi_\tau
\frac{\sqrt3 J}{4}
\sigma_y,
\label{eq:SM_Kitaev_vX}
\\
v_Y^\tau
\equiv
\partial_{k_Y}h_\tau
&=
\frac{\sqrt3}{2}
\partial_{\kappa_x}h_\tau
=
\frac{\sqrt3 J}{4}
\sigma_x.
\label{eq:SM_Kitaev_vY}
\end{align}
Combining these expressions with
Eqs.~\eqref{eq:SM_Kitaev_RB0} and
\eqref{eq:SM_Kitaev_RC0}, we obtain
\begin{align}
\begin{pmatrix}
R_{{\rm LF},B,\tau}^{(0)}
\\
R_{{\rm LF},C,\tau}^{(0)}
\end{pmatrix}
=
\frac{1}{\sqrt3}
\begin{pmatrix}
\chi_\tau&0
\\
0&-\chi_\tau
\end{pmatrix}
\begin{pmatrix}
v_X^\tau
\\
v_Y^\tau
\end{pmatrix}.
\label{eq:SM_Kitaev_Raman_velocity_operator}
\end{align}
Taking interband matrix elements therefore gives
\begin{align}
\begin{pmatrix}
M_{B,\tau}^{(0)}
\\
M_{C,\tau}^{(0)}
\end{pmatrix}
=
\frac{1}{\sqrt3}
\begin{pmatrix}
\chi_\tau&0
\\
0&-\chi_\tau
\end{pmatrix}
\begin{pmatrix}
V_{X,\tau}^{(0)}
\\
V_{Y,\tau}^{(0)}
\end{pmatrix}.
\label{eq:SM_Kitaev_Raman_velocity}
\end{align}
Here
\begin{align}
M_{P,\tau}^{(0)} &= \langle u_{+,\tau} | R_{{\rm LF},P,\tau}^{(0)} | u_{-,\tau} \rangle, \qquad P=B,C, \\
V_{\alpha,\tau}^{(0)} &= \langle u_{+,\tau} | v_\alpha^\tau | u_{-,\tau} \rangle, \qquad \alpha=X,Y,
\end{align}
are defined with respect to the same oriented physical Cartesian frame used for the Raman polarization axes. Thus both valleys realize the general relation of the main text with
\begin{align}
\lambda_K = \lambda_{K'} = \frac{1}{\sqrt3}, \qquad O_\tau =
\begin{pmatrix}
\chi_\tau&0
\\
0&-\chi_\tau
\end{pmatrix},
\qquad
\det O_\tau=-1.
\label{eq:SM_Kitaev_lambda_O}
\end{align}
Because the field-induced vertices begin only at $O(\kappa)$, the leading relation Eq.~\eqref{eq:SM_Kitaev_Raman_velocity} is independent of $\zeta_R$.

\suppsubsection{Majorana pair creation and Chern-number Raman response}
\label{sec:SM_Kitaev_pair_creation}

\begin{figure*}[t]
\centering
\includegraphics[width=\linewidth]{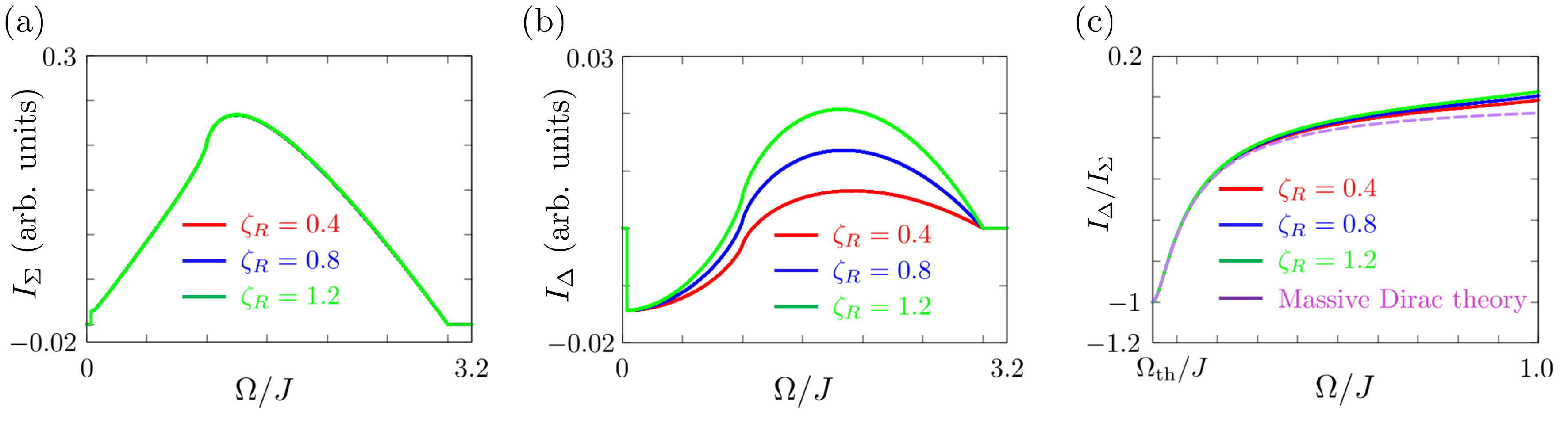}
\caption{Full-lattice Majorana Raman response for representative values of the relative field-induced Raman coupling $\zeta_R$. (a) Counter-circular sum $I_\Sigma=I_{RL}+I_{LR}$ and (b) difference $I_\Delta=I_{RL}-I_{LR}$. The individual spectra change with $\zeta_R$ because the field-induced Raman vertex modifies the matrix elements away from the Dirac points. (c) Normalized counter-circular asymmetry $I_\Delta/I_\Sigma$. All curves approach the universal threshold value $-C=-1$, independently of $\zeta_R$, because $R_{\eta,B}^{\rm Maj}$ and $R_{\eta,C}^{\rm Maj}$ vanish at the valley centers. The leading massive-Dirac prediction, $-2\Omega_{\rm th}\Omega/(\Omega^2+\Omega_{\rm th}^2)$, is shown for comparison. We use $\eta/J=0.0075$, for which $\Omega_{\rm th}/J\simeq0.039$ and $C=+1$.}
\label{fig:SM_Kitaev_zeta}
\end{figure*}

We next clarify how the two-band representation above encodes the physical Majorana Raman process.

Let $|u_\pm(\bm k)\rangle$ be the normalized eigenstates of $h_{\rm Maj}(\bm k)$ and introduce the corresponding band operators through
\begin{align}
\begin{pmatrix}
\eta_{+,\bm k} \\
\eta_{-,\bm k}
\end{pmatrix}
=
U^\dagger(\bm k)\bm\gamma_{\bm k}, \qquad U(\bm k) =
\left(|u_+(\bm k)\rangle, |u_-(\bm k)\rangle\right).
\label{eq:SM_Kitaev_band_operators}
\end{align}
Because the original fields are Majorana operators, Eq.~\eqref{eq:SM_Kitaev_PH} relates the negative-energy state at $\bm k$ to the positive-energy state at $-\bm k$. With a Majorana-compatible choice of eigenstate phases, one may take
\begin{align}
|u_-(\bm k)\rangle
=
|u_+(-\bm k)\rangle^{*},
\label{eq:SM_Kitaev_eigenstate_PH}
\end{align}
which implies
\begin{align}
\eta_{-,\bm k}
=
\eta_{+,-\bm k}^{\dagger}.
\label{eq:SM_Kitaev_band_PH}
\end{align}

After transforming to the eigenbasis of Eq.~\eqref{eq:SM_Kitaev_Raman_operator_band}, its off-diagonal contribution contains
\begin{align}
\hat R_P^{\rm off} = \frac12 \sum_{\bm k} \left[M_P(\bm k)\eta_{+,\bm k}^{\dagger}\eta_{-,\bm k} + {\rm H.c.}\right], \qquad P=B,C,
\label{eq:SM_Kitaev_Raman_offdiag}
\end{align}
where
\begin{align}
M_P(\bm k) = \langle u_+(\bm k)| R_P^{\rm Maj}(\bm k)|u_-(\bm k)\rangle, \qquad M_P(-\bm k)=-M_P(\bm k),
\label{eq:SM_Kitaev_interband_matrix_element}
\end{align}
and we use $\left[R_P^{\rm Maj}(\bm k)\right]^{T}=-R_P^{\rm Maj}(-\bm k)$. Using Eq.~\eqref{eq:SM_Kitaev_band_PH}, the positive-frequency part of Eq.~\eqref{eq:SM_Kitaev_Raman_offdiag} becomes
\begin{align}
\hat R_{P,+}^{\rm off} = \frac12 \sum_{\bm k} M_P(\bm k) \eta_{+,\bm k}^{\dagger} \eta_{+,-\bm k}^{\dagger}.
\label{eq:SM_Kitaev_pair_creation_operator}
\end{align}
Thus, the physical Raman process creates a pair of positive-energy Majorana quasiparticles with opposite momenta, while its amplitude is precisely the negative-to-positive-energy interband matrix element of the Raman vertex between the two Bloch eigenstates.
The factors of $1/2$ associated with the Majorana representation and with the redundant summation over $\bm k$ and $-\bm k$ depend on the chosen momentum-domain convention. They constitute a common polarization-independent normalization and are absorbed into ${\cal N}_R$ throughout this work. The zero-temperature positive-frequency Raman spectrum can therefore be written as
\begin{align}
I_P(\Omega)
=
{\cal N}_R
\int_{\bm k}
\left|
\langle
u_+(\bm k)
|
R_P^{\rm Maj}(\bm k)
|
u_-(\bm k)
\rangle
\right|^2
\delta
\left[
\Omega-2E(\bm k)
\right].
\label{eq:SM_Kitaev_Raman_spectrum}
\end{align}

\suppsubsection{Full-lattice response and robustness against the field-induced Raman coupling}
\label{sec:SM_Kitaev_numerics}

We finally examine the dependence of the full-lattice Raman response on the field-induced Raman coupling $\zeta_R$, using $\eta/J=0.0075$, for which $\Omega_{\rm th}/J\simeq0.039$ and $C=+1$. Figure~\ref{fig:SM_Kitaev_zeta} shows that, although the individual spectra depend on $\zeta_R$, the normalized asymmetry $I_\Delta/I_\Sigma$ closely follows the massive-Dirac prediction near the threshold and approaches the universal value $-C=-1$ for all $\zeta_R$ considered. This robustness follows from the vanishing of the field-induced Raman vertices at the Dirac points [Eq.~\eqref{eq:SM_Kitaev_Reta_valley_zero}], while deviations at higher frequencies reflect momentum-dependent lattice corrections.

\setcounter{equation}{0}

\suppsection{Topological-magnon model}
\label{sec:SM_magnon}

We consider the number-conserving honeycomb ferromagnet used in the main text. Within linear spin-wave theory, it is described by a Hermitian two-band magnon Hamiltonian. We derive its full-lattice correspondence with the Kitaev Majorana Hamiltonian and Raman vertices, and then show that the thermal occupation factor cancels exactly from the normalized counter-circular Raman response in this minimal model.

\suppsubsection{Linear-spin-wave theory and effective magnon Hamiltonian}
\label{sec:SM_magnon_H}

\begin{figure}[t]
\centering
\includegraphics[width=0.3\linewidth]{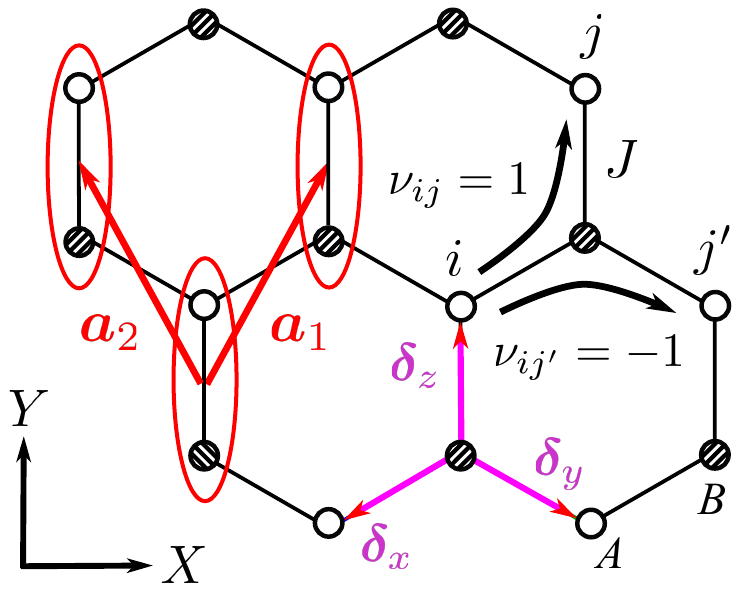}
\caption{Honeycomb-lattice convention for the topological-magnon model. Nearest-neighbor sites are coupled by ferromagnetic exchange $J$, while the arrows specify the orientation $\nu_{ij}=\pm1$ of the next-nearest-neighbor Dzyaloshinskii--Moriya interaction. The sublattices $A,B$, primitive vectors $\bm a_1,\bm a_2$, and Cartesian axes $(X,Y)$ follow the Kitaev-Majorana convention.}
\label{fig:SM_magnon_lattice}
\end{figure}

We start from the spin Hamiltonian
\begin{align}
\hat H_{\rm FM}
={}&
-J\sum_{\langle ij\rangle}\bm S_i\cdot\bm S_j
-D\sum_{\langle\!\langle ij\rangle\!\rangle}
\nu_{ij}\hat{\bm z}\cdot(\bm S_i\times\bm S_j)
-B_z\sum_i S_i^z,
\label{eq:SM_magnon_spin_H}
\end{align}
where $J>0$ is the nearest-neighbor ferromagnetic exchange, $D$ is the next-nearest-neighbor Dzyaloshinskii--Moriya interaction, and $\nu_{ij}=-\nu_{ji}=\pm1$ specifies its orientation. The lattice and DMI conventions are shown in Fig.~\ref{fig:SM_magnon_lattice}.

We expand about the collinear ferromagnetic state polarized along $+\hat{\bm z}$. Introducing a Holstein--Primakoff boson $c_i$,
\begin{align}
S_i^z&=S-c_i^\dagger c_i, &
S_i^+&=\sqrt{2S}\,c_i+O(S^{-1/2}), &
S_i^-&=\sqrt{2S}\,c_i^\dagger+O(S^{-1/2}).
\label{eq:SM_magnon_HP}
\end{align}
For a nearest-neighbor bond,
\begin{align}
\bm S_i\cdot\bm S_j
&=S_i^zS_j^z+\frac12(S_i^+S_j^-+S_i^-S_j^+)\nonumber\\
&=S^2-S(c_i^\dagger c_i+c_j^\dagger c_j)
 +S(c_i^\dagger c_j+c_j^\dagger c_i)+O(c^4).
\label{eq:SM_magnon_Heisenberg_HP}
\end{align}
As each site has three nearest neighbors, this gives
\begin{align}
-J\sum_{\langle ij\rangle}\bm S_i\cdot\bm S_j
=E_J+3JS\sum_i c_i^\dagger c_i
-JS\sum_{\langle ij\rangle}(c_i^\dagger c_j+{\rm H.c.}),
\label{eq:SM_magnon_HJ_quad}
\end{align}
where $E_J$ is a constant. For the DMI term,
\begin{align}
\hat{\bm z}\cdot(\bm S_i\times\bm S_j)=\frac{S_i^-S_j^+-S_i^+S_j^-}{2i}=-iS(c_i^\dagger c_j-c_j^\dagger c_i)+O(c^4).
\label{eq:SM_magnon_DMI_HP}
\end{align}
Consequently,
\begin{align}
-D\sum_{\langle\!\langle ij\rangle\!\rangle} \nu_{ij}\hat{\bm z}\cdot(\bm S_i\times\bm S_j) =-DS\sum_{\langle\!\langle ij\rangle\!\rangle} \bigl(i\nu_{ij}c_j^\dagger c_i+{\rm H.c.}\bigr).
\label{eq:SM_magnon_HD_quad}
\end{align}
Reversing all the $\nu_{ij}$ is equivalent to reversing the sign of $D$. The Zeeman term becomes
\begin{align}
-B_z\sum_i S_i^z=E_B+B_z\sum_i c_i^\dagger c_i,
\label{eq:SM_magnon_HB_quad}
\end{align}
where $E_B$ is constant. These quadratic terms conserve magnon number: there are no anomalous terms $c_ic_j$ or $c_i^\dagger c_j^\dagger$, so a bosonic Bogoliubov treatment is unnecessary.

We use the unit-cell and primitive-vector convention of Eq.~\eqref{eq:SM_Kitaev_lattice_vectors} and define $p_i=\bm k\cdot\bm a_i$. The Fourier transform of the original Holstein--Primakoff bosons is $c_{\bm r,s}=N^{-1/2}\sum_{\bm k}e^{i\bm k\cdot\bm r}c_{\bm k,s}$, with $s=A,B$. The nearest-neighbor matrix element is $-JSf(\bm k)$, where
\begin{align}
f(\bm k)=1+e^{-ip_1}+e^{-ip_2}.
\label{eq:SM_magnon_structure_factor}
\end{align}
For direct comparison with the Kitaev Majorana Hamiltonian, we make the momentum-independent rephasing $b_{\bm k}=i c_{\bm k,B}$, leaving $a_{\bm k}=c_{\bm k,A}$ unchanged, and use the spinor
\begin{align}
\bm b_{\bm k}= \begin{pmatrix}a_{\bm k}\\b_{\bm k}\end{pmatrix}.
\label{eq:SM_magnon_spinor}
\end{align}
In this basis, the nearest-neighbor term is
\begin{align}
d_x^{\rm M}(\bm k)+id_y^{\rm M}(\bm k) = -iJS\bigl(1+e^{-ip_1}+e^{-ip_2}\bigr).
\label{eq:SM_magnon_dxy}
\end{align}
The DMI hopping connects sites on the same sublattice and is therefore unchanged by this rephasing. With the orientations of Fig.~\ref{fig:SM_magnon_lattice}, its diagonal contribution is
\begin{align}
d_z^{\rm M}(\bm k) = 2DS \bigl[-\sin p_1+\sin p_2+\sin(p_1-p_2)\bigr].
\label{eq:SM_magnon_dz}
\end{align}
The factor of two comes from adding each oriented hopping and its Hermitian conjugate.

Dropping the constant $E_0=E_J+E_B$, we obtain
\begin{align}
\hat H_{\rm M}&=\sum_{\bm k}\bm b_{\bm k}^\dagger
h_{\rm M}(\bm k)\bm b_{\bm k},
\label{eq:SM_magnon_Bloch_H}\\
h_{\rm M}(\bm k)&=\varepsilon_{\rm D}\sigma_0
+\sum_{a=x,y,z}d_a^{\rm M}(\bm k)\sigma_a,
\label{eq:SM_magnon_h}\\
\varepsilon_{\rm D}&=3JS+B_z.
\label{eq:SM_magnon_epsD}
\end{align}
The two magnon energies and their direct interband separation are
\begin{align}
\varepsilon_\pm(\bm k)&=\varepsilon_{\rm D}\pm E(\bm k),
\qquad E(\bm k)=\sqrt{\sum_a[d_a^{\rm M}(\bm k)]^2},
\label{eq:SM_magnon_energies}\\
\Delta_{\rm M}(\bm k)&=\varepsilon_+(\bm k)-\varepsilon_-(\bm k)
=2E(\bm k).
\label{eq:SM_magnon_transition_energy}
\end{align}
The momentum-independent term $\varepsilon_{\rm D}\sigma_0$ changes neither the eigenstates nor their quantum geometry and cancels from $\Delta_{\rm M}$.

\suppsubsection{Magnon Raman vertices and mapping to the Majorana model}
\label{sec:SM_magnon_Raman}

The nearest-neighbor exchange contribution to the Raman operator is of Loudon--Fleury form,
\begin{align}
\hat R_J=-J\sum_{\langle ij\rangle}
(\bm e_{\rm in}\cdot\bm\delta_{ij})
(\bm e_{\rm out}^{*}\cdot\bm\delta_{ij})
\bm S_i\cdot\bm S_j.
\label{eq:SM_magnon_LF}
\end{align}
Using Eq.~\eqref{eq:SM_Kitaev_bond_vectors} and the same rephased boson basis as above, its symmetric-traceless components are
\begin{align}
R_{J,P}(\bm k)&=
\begin{pmatrix}0&z_{J,P}^{*}(\bm k)\\z_{J,P}(\bm k)&0\end{pmatrix},
\qquad P=B,C,
\label{eq:SM_magnon_RJ_matrix}\\
z_{J,B}(\bm k)&=-\frac{iJS}{6}
\bigl(e^{-ip_1}+e^{-ip_2}-2\bigr),
\label{eq:SM_magnon_RJ_B}\\
z_{J,C}(\bm k)&=-\frac{iJS}{2\sqrt3}
\bigl(e^{-ip_1}-e^{-ip_2}\bigr).
\label{eq:SM_magnon_RJ_C}
\end{align}

For the DMI contribution, we use the effective LMC prescription of Sec.~\ref{sec:SM_LMC_vertex}, with the same sign convention as in the Kitaev Majorana model:
\begin{align}
R_{D,\alpha\beta}(\bm k) =-\partial_{k_\alpha}\partial_{k_\beta} d_z^{\rm M}(\bm k)\,\sigma_z, \qquad \alpha,\beta=X,Y.
\label{eq:SM_magnon_RD_tensor}
\end{align}
It gives
\begin{align}
R_{D,B}(\bm k)&=2DS
\left[\frac12\sin p_1-\frac12\sin p_2
+\sin(p_1-p_2)\right]\sigma_z,
\label{eq:SM_magnon_RD_B}\\
R_{D,C}(\bm k)&=-\sqrt3 DS
(\sin p_1+\sin p_2)\sigma_z.
\label{eq:SM_magnon_RD_C}
\end{align}
The full Raman operators used in the calculation are
\begin{align}
\hat R_P&=\sum_{\bm k}\bm b_{\bm k}^{\dagger}
R_P^{\rm M}(\bm k)\bm b_{\bm k},
\label{eq:SM_magnon_Raman_operator_band}\\
R_P^{\rm M}(\bm k)&=R_{J,P}(\bm k)+R_{D,P}(\bm k),
\qquad P=B,C.
\label{eq:SM_magnon_full_Raman}
\end{align}

Equations~\eqref{eq:SM_magnon_dxy} and \eqref{eq:SM_magnon_dz}, together with the Raman vertices above, establish the full-lattice identities
\begin{align}
h_{\rm M}(\bm k)-\varepsilon_{\rm D}\sigma_0
&=h_{\rm Maj}(\bm k)
\big|_{J\to2JS,\,\eta\to2DS},
\label{eq:SM_magnon_Kitaev_mapping_H}\\
R_P^{\rm M}(\bm k)
&=R_P^{\rm Maj}(\bm k)
\big|_{J\to2JS,\,\eta\to2DS,\,\zeta_R\to1},
\qquad P=B,C.
\label{eq:SM_magnon_Kitaev_mapping_R}
\end{align}

Using Eq.~\eqref{eq:SM_Kitaev_Dirac} and the same physical Cartesian momentum frame, the magnon Hamiltonian near either valley is
\begin{align}
h_{{\rm M},\tau}(\bm\kappa)
=\varepsilon_{\rm D}\sigma_0
+JS\kappa_x\sigma_x
+JS\chi_\tau\kappa_y\sigma_y
+m_{{\rm M},\tau}\sigma_z+O(\kappa^2),
\label{eq:SM_magnon_Dirac}
\end{align}
with
\begin{align}
m_{{\rm M},\tau}&=-\chi_\tau3\sqrt3 DS.
\label{eq:SM_magnon_valley_mass}
\end{align}
Therefore,
\begin{align}
\Omega_{\rm th}&=2|m_{{\rm M},\tau}|=6\sqrt3|D|S,
\label{eq:SM_magnon_threshold}\\
C_{\rm M}&=\operatorname{sgn}(D)
\label{eq:SM_magnon_Chern}
\end{align}
for the lower magnon band and the orientation convention used here. The leading Raman--velocity relation also follows directly from Eq.~\eqref{eq:SM_Kitaev_Raman_velocity}, with $\lambda_K=\lambda_{K'}=1/\sqrt3$ and $\det O_\tau=-1$.

\suppsubsection{Finite-temperature interband Raman process}
\label{sec:SM_magnon_finiteT}

We next express the Raman process in the magnon band eigenbasis. Let $|u_\pm(\bm k)\rangle$ be the normalized eigenstates of $h_{\rm M}(\bm k)$, and define the magnon band operators by
\begin{align}
\begin{pmatrix}
\alpha_{+,\bm k}\\
\alpha_{-,\bm k}
\end{pmatrix}
&=U_{\rm M}^{\dagger}(\bm k)\bm b_{\bm k},\qquad
U_{\rm M}(\bm k)
=\bigl(|u_+(\bm k)\rangle,|u_-(\bm k)\rangle\bigr).
\label{eq:SM_magnon_band_operators}
\end{align}
These are independent canonical bosonic annihilation operators, $[\alpha_{s,\bm k},\alpha_{s',\bm k'}^{\dagger}] =\delta_{ss'}\delta_{\bm k\bm k'}$, with $s,s'=\pm$. Unlike the two-Majorana creation process, the number-conserving magnon Raman vertex transfers an already occupied magnon from the lower band to the upper band. For the Hermitian Cartesian components $P=B,C$, the interband operator reads
\begin{align}
\hat R_P^{\rm inter} = \sum_{\bm k}\left[ M_P(\bm k)\alpha_{+,\bm k}^\dagger\alpha_{-,\bm k} +{\rm H.c.}\right],
\label{eq:SM_magnon_interband_Raman}
\end{align}
where
\begin{align}
M_P(\bm k)=\langle u_+(\bm k)|R_P^{\rm M}(\bm k)|u_-(\bm k)\rangle.
\label{eq:SM_magnon_MP}
\end{align}
The counter-circular vertices are obtained by the linear combinations in Eq.~\eqref{eq:SM_counter_circular_vertices}. At finite temperature, the positive-frequency Stokes spectrum is
\begin{align}
\widetilde I_P(\Omega,T) ={\cal N}_R\int_{\bm k}{\cal B}_T(\bm k) |M_P(\bm k)|^2\delta[\Omega-\Delta_{\rm M}(\bm k)],
\label{eq:SM_magnon_finiteT_spectrum}
\end{align}
with the Bose occupation factor
\begin{align}
{\cal B}_T(\bm k)
=n_B[\varepsilon_-(\bm k)]
\bigl\{1+n_B[\varepsilon_+(\bm k)]\bigr\},
\qquad n_B(\epsilon)=\frac{1}{e^{\epsilon/T}-1}.
\label{eq:SM_magnon_Bose_factor}
\end{align}
The first factor counts occupied lower-band magnons; the second accounts for Bose enhancement in the final band.

Because $\varepsilon_{\rm D}$ is momentum independent, a Raman shell $\Omega=\Delta_{\rm M}(\bm k)=2E(\bm k)$ has
\begin{align}
\varepsilon_-(\bm k)&=\varepsilon_{\rm D}-\frac{\Omega}{2}, &
\varepsilon_+(\bm k)&=\varepsilon_{\rm D}+\frac{\Omega}{2}.
\label{eq:SM_magnon_shell_energies}
\end{align}
Thus the Bose factor is constant on the shell:
\begin{align}
{\cal B}_T(\Omega)
=n_B\left(\varepsilon_{\rm D}-\frac{\Omega}{2}\right)
\left[1+n_B\left(\varepsilon_{\rm D}+\frac{\Omega}{2}\right)\right].
\label{eq:SM_magnon_Bose_shell}
\end{align}
Defining $I_P(\Omega)$ by the right-hand side of Eq.~\eqref{eq:SM_magnon_finiteT_spectrum} without ${\cal B}_T$, we obtain the exact full-lattice factorization
\begin{align}
\widetilde I_P(\Omega,T)={\cal B}_T(\Omega)I_P(\Omega).
\label{eq:SM_magnon_factorization}
\end{align}
Consequently, with $I_\Sigma=I_{RL}+I_{LR}$ and $I_\Delta=I_{RL}-I_{LR}$ (and analogously for $\widetilde I$),
\begin{align}
\frac{\widetilde I_\Delta(\Omega,T)}
{\widetilde I_\Sigma(\Omega,T)}
=\frac{I_\Delta(\Omega)}{I_\Sigma(\Omega)}.
\label{eq:SM_magnon_thermal_cancel}
\end{align}
The cancellation holds at every Raman shift where the signal and ratio are defined; it does not rely on the Dirac approximation. At strictly $T=0$, the initial lower magnon band is unoccupied, so $\widetilde I_P(\Omega,0)=0$ for this interband Stokes process and the physical ratio is undefined.

Combining the exact cancellation with the leading massive-Dirac result gives, for finite $T$ and a nonzero signal,
\begin{align}
\frac{\widetilde I_\Delta(\Omega,T)}
{\widetilde I_\Sigma(\Omega,T)}
=-C_{\rm M}\frac{2\Omega_{\rm th}\Omega}
{\Omega^2+\Omega_{\rm th}^2},
\qquad
\lim_{\Omega\to\Omega_{\rm th}^+}
\frac{\widetilde I_\Delta(\Omega,T)}
{\widetilde I_\Sigma(\Omega,T)}
=-C_{\rm M}.
\end{align}

\begin{figure*}[t]
\centering
\includegraphics[width=\linewidth]{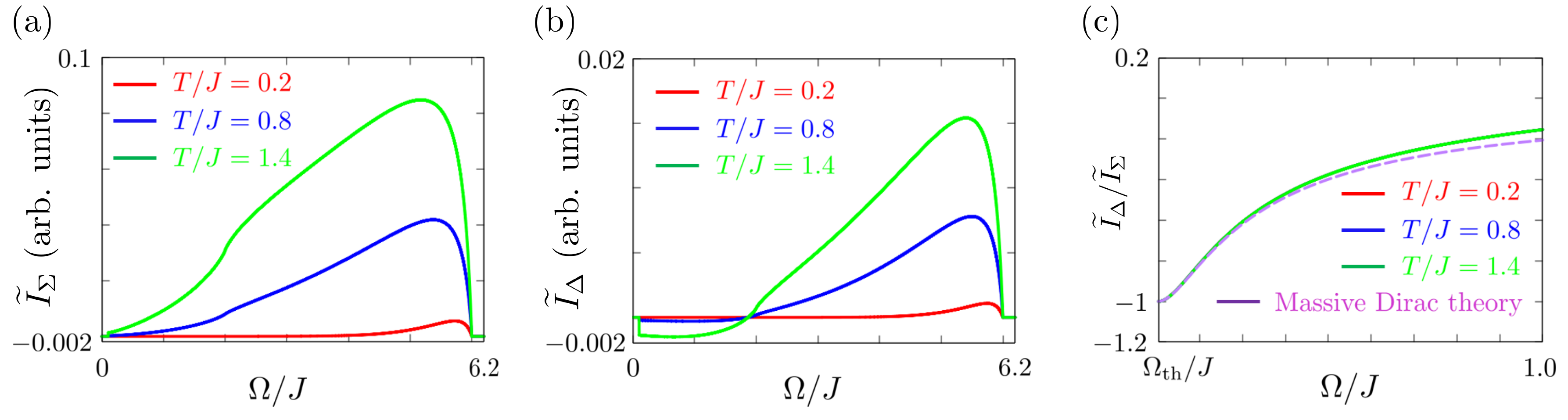}
\caption{Full-lattice finite-temperature Raman response of the number-conserving topological-magnon model. (a) Counter-circular sum $\widetilde I_\Sigma= \widetilde I_{RL}+\widetilde I_{LR}$ and (b) difference $\widetilde I_\Delta= \widetilde I_{RL}-\widetilde I_{LR}$ at representative temperatures. (c) Normalized asymmetry $\widetilde I_\Delta/\widetilde I_\Sigma$. The temperature-dependent Bose factor cancels exactly from the ratio; the massive-Dirac prediction is shown for comparison. We use $D/J=0.01$, $B_z/J=0.1$, and $S=1$, for which $\Omega_{\rm th}/J\simeq0.104$ and $C_{\rm M}=+1$.}
\label{fig:SM_magnon_temperature}
\end{figure*}

\suppsubsection{Full-lattice temperature dependence}
\label{sec:SM_magnon_numerics}

We finally examine the full-lattice finite-temperature Raman response for $D/J=0.01$, $B_z/J=0.1$, and $S=1$, using $J$ as the energy unit. These parameters give $\Omega_{\rm th}/J\simeq0.104$ and $C_{\rm M}=+1$. Figure~\ref{fig:SM_magnon_temperature} shows that the absolute spectra vary strongly with temperature, while the normalized asymmetry collapses onto the same full-lattice curve at all temperatures considered. Near the threshold it approaches $-C_{\rm M}=-1$; deviations from the massive-Dirac prediction at higher frequencies arise from lattice corrections, not thermal occupation effects.

\end{document}